\documentclass{article}

\usepackage{arxiv}

\usepackage[utf8]{inputenc} 
\usepackage[T1]{fontenc}    
\usepackage{fancyhdr}
\usepackage{lmodern}
\usepackage[colorlinks]{hyperref}  
\usepackage{url}            
\usepackage{booktabs}       
\usepackage{amsfonts}       
\usepackage{nicefrac}       
\usepackage{microtype}      
\usepackage{lipsum}

\usepackage{graphicx}
\usepackage{epsfig}
\usepackage{epstopdf}
\usepackage{subfigure}

\usepackage{enumerate}
\usepackage{amssymb}
\usepackage{amsmath}
\usepackage{amsthm}
\usepackage{enumitem}
\usepackage{mathrsfs}

\newcommand{\be}{\begin{equation}}
\newcommand{\ee}{\end{equation}}
\newcommand{\bea}{\begin{eqnarray}}
\newcommand{\eea}{\end{eqnarray}}
\newcommand{\beaa}{\begin{eqnarray*}}
\newcommand{\eeaa}{\end{eqnarray*}}

\newcommand{\eps}{\varepsilon}

\newcommand{\norm}[1]{\left\|#1\right\|}

\def\dis{\displaystyle}

\newcommand{\dsp}{\displaystyle}

\theoremstyle{thmstyleone}%
\newtheorem{theorem}{Theorem}
\theoremstyle{thmstyletwo}%

\theoremstyle{thmstylethree}%

\title{Turing patterns in a Leslie-Gower predator prey model}

\author{  
F. Capone$^a$\thanks{Corresponding author.} \\
\texttt{fcapone@unina.it} \\
\And 
R. De Luca$^a$  \\  
\texttt{roberta.deluca@unina.it} \\     
\And  
I. Torcicollo$^b$ \\ 
\texttt{i.torcicollo@iac.cnr.it}
\\ \\ 
$^a$Dipartimento di Matematica e Applicazioni "R.Caccioppoli" \\ Università degli Studi di Napoli Federico II \\ Via Cintia, Monte S.Angelo, 80126 Napoli \\ Italy  \\[2mm]
$^b$Istituto per le Applicazioni del Calcolo "Mauro Picone", CNR \\ Via Pietro Castellino, 80129 Napoli \\ Italy}

\renewcommand{\shorttitle}{Turing patterns in a Leslie-Gower predator prey model}

\begin{document}
\maketitle

\begin{abstract}
A reaction-diffusion Leslie-Gower predator-prey model, incorporating the fear effect and prey refuge, with Beddington-DeAngelis functional response, is introduced. A qualitative analysis of the solutions of the model and the stability analysis of the coexistence equilibrium, are performed. Sufficient conditions guaranteeing the occurrence of Turing instability have been determined either in the case of self-diffusion or in the case of cross-diffusion. Different types of Turing patterns, representing a spatial redistribution of population in the environment, emerge for different values of the model parameters.  
\end{abstract}

\section{Introduction}
Predator-prey models describe the interaction between two population in which a species (the predators) sustains its self by eating another one (the prey). Starting from the pioneering Lotka-Volterra predator-prey model, different generalizations have been proposed in order to overcome some criticalities and better describe some real situations \cite{CDR,CCDTIntraguild,CCDTTuring,CCDT,CT,J2,D,J1}, \cite{Murray1,Murray2}. In particular, a Leslie-Gower model has been successively formulated in order to introduce an asymptotic limit to the growth of both populations (not recognized by the classical model) \cite{Chen,Y,GC,K}. This model consists in two ordinary differential equations in which the environmental carrying capacity of predators depends on the ratio between the two population densities. A fundamental role in mathematical modeling of predator-prey dynamics, is played by the functional response defined as the number of prey consumed by one predator per unit of time. The functional response depends on a number of factors such as the prey’s ability to escape an attack or the predator’s search efficiency. Holling \cite{H} proposed three functional responses depending only on the number of prey ($N$).
In particular, the Holling functionals are:
\begin{itemize}
\item[$\bullet$] Type I: $\mathcal F=mN$;

\item[$\bullet$] Type II: $\mathcal F=\dis\frac{mN}{a+N}$;

\item[$\bullet$] Type III: $\mathcal F=\dis\frac{mN^2}{a+N^2}$
\end{itemize}
being $a,m$ constants.
The choice of the functional response depends on 
the different predation behaviour to be modeled. In particular, the type I is used when there is no handling time
of the captured prey and population densities are not too large. Type II introduces a maximum predation rate to describe the situation in which predators can feel satiated when there is abundant available food. Type III describes the increasing of predators search activity  with increasing prey density. 
However, there are some circumstances  in which a functional response, depending of both population densities, should be used. This is the case, for example, in which predators behavior affects the prey dynamics. Beddington-DeAngelis \cite{B,BDA} proposed a functional response which comes from a generalization of the Holling type II functional response, introducing at the denominator an additive linear term depending on the predators number to model the mutual interference between predators.  In \cite{F} a modified Leslie-Gower model has been introduced to describe the predator-prey interaction by considering a Beddington-DeAngelis functional response and taking into account of two important aspects: the fear effect and the prey refuge.  Fear may have important consequences on the ecosystem \cite{Cr,Wang,Fear2,Fear3}. For example, in \cite{Cr} it has been observed that some birds female, that experienced frequent nests predation, produce fewer eggs in the subsequent nests. In order to model this phenomenon, the natural birth rate of preys is multiplied by a function $f(k,P)$ depending on the level of fear $k$ and on the predators density $P$ such as:
\begin{itemize}
\item[1)]	in the absence of fear or in the absence of predators, the function is equal to 1, meaning that the natural birth rate of preys is constant
$$    f(0,P)=1, \qquad f(k,0)=1 $$
\item[2)]	when the level of fear or the predators density is huge, the function $f$ tends to zero
$$
\lim_{k\to\infty} f(k,P)=0,\qquad \lim_{P\to\infty} f(k,P)=0
$$
\item[3)]	the function $f$ has to be decreasing with respect to $k$ and $P$
$$
\frac{\partial f(k,P)}{\partial k}<0,\qquad \frac{\partial f(k,P)}{\partial P}<0.
$$
\end{itemize}

In time of predation, preys can experience hiding behavior
\cite{Gosh}, \cite{Sih}, \cite{FearRefuge}. Then, introducing a parameter $\eta\in[0,1]$ representing the fraction of prey protected by predation, denoting by $N$ the number of prey, $\eta(1-N) $ is the number of prey outside of protection. The model introduced in \cite{F} considers the
case in which both populations are homogeneously mixed in the environment. The biologically meaningful equilibria have been determined and sufficient conditions guaranteeing the linear stability of the coexistence equilibrium have been found. \\
\noindent In order to generalize the results obtained in \cite{F}, in this paper we consider the case in which populations are heterogeneously mixed in the environment to incorporate a random movement of both species. Such model better describes, for example, the situations in which predators can move to search for preys and these ones can move to escape by predators attack. When diffusion is incorporated in the mathematical model, a spatial distribution, periodic in space and constant in time, of both populations can be observed under certain conditions (see \cite{CC} and the references therein). In fact, it is possible to look for conditions guaranteeing that an equilibrium, stable in the absence of diffusion, becomes unstable when diffusion is allowed. The diffusion-driven instability is called Turing instability and has been widely studied in literature, especially to investigate for the Turing patterns formation (\cite{Turing}). This approach can be extended to other interacting models with different functional responses,
and also in other fields of applied mathematics where nonlinear mathematical models having a similar structure are considered (\cite{T,RT,CTi,JN}). The simplest diffusion is the linear one, meaning that the time evolution of one species is affected by the random movement of the same species.
In \cite{CZ} a modified Leslie-Gower model is introduced. It is assumed a linear constant self diffusion and conditions guaranteeing the onset of Turing, Hopf, Turing-Hopf bifurcations, is investigated.
However, more sophisticated diffusion terms can be introduced due to the fact that
 the interaction between population affects each other's behaviour. Among these, the cross-diffusion terms are introduced when the behaviour of one species depends on the random movements of both species.\\
\noindent
The plan of the paper is as follows. Section 2 is devoted to the introduction of the mathematical model which consists of two reaction-diffusion equations governing the evolution of predators and prey interactions. A simple linear, constant, self-diffusion is introduced for both the species. Section 3 is devoted to a qualitative analysis of the solutions: the boundedness and existence of absorbing sets (i.e. positively invariant and attractive sets) in the phase space are explored.  In the subsequent Section 4, the existence of biologically meaningful equilibria is analyzed. Section 5 deals with the linear instability of the coexistence equilibrium. Precisely, in Subsecion 5.1, the linear instability in the homogeneous case is investigated. The heterogeneous case is examined in Subsection 5.2 where
sufficient conditions guaranteeing the occurrence of Turing instability, have been determined. Since the set of parameters verifying the conditions for the diffusion-driven instability is very strict, in Section 6 the model introduced in Section 2, has been generalized to take into account of cross-diffusion and conditions guaranteeing the onset of Turing instability have been determined in the case in which this kind of instability is not observable when the self-diffusion is considered. In Section 7 the amplitude equations are obtained. Section 8 deals with numerical simulations in order to explore a richer dynamic of population interactions showing that, under certain conditions, spatial patterns emerge. The paper ends with a Conclusion section
(Section 9) collecting  all the obtained results.

\section{Mathematical model}\label{model}
In \cite{F}, a Leslie-Gower predator-prey model with Beddington-DeAngelis functional response, incorporating fear effect and prey refuge, has been analyzed.
Denoting by $N$ and $P$ the number of prey and predators, respectively, the model introduced in \cite{F} is 
\be\label{1}
\begin{cases}
 \dis\frac{dN}{dt}=\left(\dis\frac{r_1}{1+kP}-qN-\dis\frac{\alpha(1-\eta)P}{a+b(1-\eta)N+cP}-d\right)N,\\
 \dis\frac{dP}{dt}=\left(r_2-\dis\frac{\beta P}{(1-\eta)N+\gamma}\right)P
\end{cases}\ee
with $r_1,r_2$ birth rates of prey and predator population, $q, \beta$ competition rates of prey and predators; $\alpha$ reduction rate of prey into predators, $a,\gamma$ environmental protects of prey and predators, $b,c$ constants, $d$ natural death rate of prey, $\eta\in[0,1[$ measures the protection of prey (i.e. $\eta$ is the number of prey protected by predation), $k$ rate of fear expressing the anti-predator behaviour in prey. All the constants appearing in (\ref{1}) are positive.
In model (\ref{1}), population is considered homogeneously mixed in the environment (i.e. diffusion is neglected). In order to generalize model (\ref{1}) to the most significant case in which both species can randomly move in the environment (for example, prey can escape from regions with high risk of predation, or predators can move for searching food), we introduce -- at the first -- the simplest diffusion, i.e. the linear, constant self-diffusion terms, to obtain
\be\label{mod}
\begin{cases}
 \dis\frac{\partial N}{\partial \tau}=\left(\dis\frac{r_1}{1+kP}-qN-\dis\frac{\alpha(1-\eta)P}{a+b(1-\eta)N+cP}-d\right)N+d_1\Delta N,\\
 \dis\frac{\partial P}{\partial \tau}=\left(r_2-\dis\frac{\beta P}{(1-\eta)N+\gamma}\right)P+d_2\Delta P
\end{cases}\ee
where $d_i$ are positive constants ($i=1,2$), denoting the diffusion coefficients and $\Delta$ is the spatial Laplacian operator. 
In the sequel, we denote by $\Omega$ the domain in which species can spread, assuming that $\Omega$ is a regular domain, and associate to (\ref{mod}) smooth positive initial data:
\be\label{cin}
N(\mathbf X, 0)=N_0(\mathbf X),\qquad P(\mathbf X,0)=P_0(\mathbf X),\qquad \mathbf X\in\Omega\ee
and homogeneous Neumann boundary conditions (no-flux)
\be\label{cbor}
\nabla N\cdot\mathbf n=0,\qquad \nabla P\cdot\mathbf n=0,\qquad \mbox{on}\,\, \partial\Omega\times\mathbb R^+,\ee
being $\mathbf n$ the outward unit normal to the boundary $\partial\Omega$.
Introducing the transformation (see \cite{F})
\be\label{tra}
\begin{cases}
 \mathbf x=\dis\frac{\mathbf X}{L}, \,\,\, t=r_2\tau,\,\,\, n=\dis\frac{qN}{r_2},\,\,\,p=\dis\frac{qP}{br^2_2},\,\,\,
 \mu=\dis\frac{r_1}{r_2},\,\,\, \rho=\dis\frac{kbr^2_2}{q},\\
 \delta=\dis\frac{q}{r_2},\,\,\,\sigma=\dis\frac{b r_2}{a q},\,\,\,
 \Phi=\dis\frac{\alpha b r_2}{a q },\,\,\,\xi=\dis\frac{b c r^2_2}{a q },\,\,\,\theta=b \beta,\,\,\,\nu=\dis\frac{\gamma q}{r_2}
\end{cases}\ee
with $L$ being the $\Omega$-diameter, setting $\gamma_i=\dis\frac{d_i}{r_i L^2},\,(i=1,2)$, model (\ref{mod}) becomes
\be\label{modadi}
\begin{cases}
 \dis\frac{\partial n}{\partial t}=\left(\dis\frac{\mu}{1+\rho p}-n-\dis\frac{(1-\eta)\Phi p}{1+\sigma(1-\eta)n+\xi p}-\delta\right)n+\gamma_1\Delta n,\\
 \dis\frac{\partial p}{\partial t}=\left(1-\dis\frac{\theta}{(1-\eta)n+\nu}\right)p+\gamma_2\Delta p
\end{cases}
\ee
under the initial-boundary conditions
\be\label{ibc}
\begin{array}{l}
 n(\mathbf x,0)=\mathbf n_0(\mathbf x),\quad  p(\mathbf x,0)=p_0(\mathbf x),\quad \mathbf x\in \Omega,\\
 \\
 \nabla n\cdot\mathbf n=0,\quad \nabla p\cdot\mathbf n=0,\quad \mbox{on }\partial\Omega\times\mathbb R^+.
\end{array}\ee
In the sequel we assume, accordingly to \cite{F}, that $\mu>\delta$.

\section{Boundedness of solutions}
In this section, we investigate the boundedness of solutions and the existence of absorbing sets in the phase space (i.e. positively invariant and attractive sets).
Denote by $\|\cdot\|, \|\cdot\|_\infty$ the $L^2$ and $L^\infty-$norm. Let $T>0$ be a fixed time and $\Omega_T=\Omega\times(0,T]$ be the parabolic cylinder. The following theorem holds true.
\begin{theorem}
 Let $(n,p)\in [C^2_1(\Omega_T)\cap C(\bar\Omega_T)]^2$ the non negative solution of (\ref{modadi})-(\ref{ibc}).
Then $\forall \varphi\in\{n,p\}$,
 $\varphi$ is bounded a.e. in $\Omega$ according to
\be\label{bound}
 n(\mathbf x, t)\leq C_\infty^{(1)}(n_0(\mathbf x)):=M_1,\quad
 p(\mathbf x, t)\leq C_\infty^{(2)}({p_0(\mathbf x)}):=M_2,
\ee
where $C_\infty^{i}$, ($i=1,2$) are positive constants depending on the initial data.
\end{theorem}
\textbf{Proof.}
$n(\mathbf x, t)$ is a sub-solution of the problem
\be\label{subs}
\begin{cases}
 \dis\frac{\partial S_1}{\partial t}-\gamma_1\Delta S_1=(\mu-\delta-S_1) S_1,\\
 \nabla S_1\cdot\mathbf n=0,\quad\mbox{ on }\partial\Omega\times\mathbb R^+,\\
 S_1(\mathbf x, 0)=S^0_1(\mathbf x)=\dis\max_{\bar\Omega}n_0(\mathbf x).
\end{cases}\ee
Since
\be\label{pog}
(\mu-\delta-S_1)S_1\leq \dis\frac{3}{2} S_1^2+\dis\frac{(\mu-\delta)^2}{2},\ee
in view of Theorem 1 of \cite{Le},
one obtains that, denoting by $\tau_1(S^0_1)$ the maximal existence time of the solution $S_1(\mathbf x, t)$
of (\ref{subs}),
since -- from the continuous dependence on the initial data -- 
there exists a positive  constant $C_1(S^0_1)$ such that
\be\norm{S_1(\cdot,t)}\leq C_1(S^0_1),\,\forall t\in (0,\tau_1 (S^0_1)),\ee
the solution $S_1(\mathbf x, t)$ exists for all time and there exists a positive constant $C_{\infty}^{(1)}$ such
that 
\be\label{dis}
\norm{S_1(\cdot,t)}_{\infty}\leq C_{\infty}^{(1)}(S^0_1),\,\forall t>0.\ee
Similarly, 
$p(\mathbf x, t)$ is a sub-solution of the problem
\be\label{subs2}
\begin{cases}
 \dis\frac{\partial S_2}{\partial t}-\gamma_2\Delta S_2=\left(1-\dis\frac{\theta}{(1-\eta)M_1+\nu}\right) S_2,\\
 \nabla S_2\cdot\mathbf n=0,\quad\mbox{ on }\partial\Omega\times\mathbb R^+,\\
 S_2(\mathbf x, 0)=S^0_2(\mathbf x)=\dis\max_{\bar\Omega}p_0(\mathbf x).
\end{cases}\ee
Since
\be\label{pog2}
\left(1-\dis\frac{\theta}{(1-\eta)M_1+\nu}\right) S_2\leq \left(1-\dis\frac{\theta}{(1-\eta)M_1+\nu}\right)^2 \dis\frac{S_2^2}{2}+\dis\frac{1}{2},\ee
in view of Theorem 1 of \cite{Le},
one obtains that, denoting by $\tau_2(S^0_2)$ the maximal existence time of the solution $S_2(\mathbf x, t)$
of (\ref{subs2}),
since -- from the continuous dependence on the initial data -- 
there exists a positive  constant $C_2(S^0_2)$ such that
\be\norm{S_2(\cdot,t)}\leq C_2(S^0_2),\,\forall t\in (0,\tau_2 (S^0_2)),\ee
the solution $S_2(\mathbf x, t)$ exists for all time and there exists a positive constant $C_{\infty}^{(2)}$ such
that 
\be\label{dis2}
\norm{S_2(\cdot,t)}_{\infty}\leq C_{\infty}^{(2)}(S^0_2),\,\forall t>0
\ee
and the thesis is proved.
\begin{theorem}
$\forall\eps>0$ the manifold
\be\label{3.8}
\Sigma_\eps\!\!=
\left\{(n,p)\!\in[\mathbf R^{+}]^2\!:\!\left\| n\right\|^2\!+\!\left\| p\right\|^2\leq(1+\eps)\dis\frac{\bar a}{\delta}\right\}
\ee
 with
\be\label{3.9}
\bar{a}=2\vert\Omega\vert\left[(\sigma+1){M^2}_1+M_1 C^2_\infty(1+\alpha C_{\infty})\right],
\ee
is an absorbing set for system (\ref{modadi}).
\end{theorem}
\textbf{Proof.} Multiplying (\ref{modadi})$_1$ by $n$, (\ref{modadi})$_2$ by $p$, adding the resulting equations and integrating
over $\Omega$, by virtue of the divergence theorem, the boundary conditions (\ref{ibc})$_2$ and 
(\ref{bound}), it turns out that
\be
\dis\frac{1}{2}\dis\frac{d(\norm{n}^2+\norm{p}^2)}{dt} \leq -\delta(\|n\|^2+\|p\|^2)+\tilde a,\ee
with 
$ \tilde a=\left(M^3_1+(1-\eta)\Phi M^2_1 M_2 +\mu M^2_1 \beta +\dis\frac{\theta}{2} M^2_2+(1+\delta)M^2_2\right)\vert\Omega\vert$.
Then, setting $E=\norm{n}^2+\norm{p}^2$, it follows that
\be
\dis\frac{dE}{dt}\leq -2\delta E+2\bar a.\ee
Following the procedure in \cite{sal}, one can prove that $\Sigma$ is an absorbing set.
\section{Biologically meaningful equilibria: existence and a priori estimates}
The biologically meaningful equilibria are the positive solutions of the system
\be\label{equi}
\begin{cases} \left(\dis\frac{\mu}{1+\rho p}-n-\dis\frac{(1-\eta)\Phi p}{1+\sigma(1-\eta)n+\xi p}-\delta\right)n+\gamma_1\Delta n=0,\\
\left(1-\dis\frac{\theta}{(1-\eta)n+\nu}\right)p+\gamma_2\Delta p=0.
\end{cases}\ee
The following Theorem holds true.
\begin{theorem}
 Let $0<\gamma\leq\dis\min\{\gamma_1,\gamma_2\}$. Then there exist positive constants 
 $C_i(\gamma)\,(i=1,2)$ depending on the positive constants appearing in (\ref{modadi}) and $\Omega$ such that any positive solution
 of (\ref{equi}) verifies:
 \be\label{bouneq}
 \begin{array}{l}
  \dis\max_{\bar\Omega}n(\mathbf x)\leq M_1,
  \quad \dis\max_{\bar\Omega} p(\mathbf x)\leq M_2,\\
  \\
  \dis\frac{\max_{\bar\Omega}n(\mathbf x)}{\min_{\bar\Omega} n(\mathbf x)}\leq C_1(\gamma),\quad
  \dis\frac{\max_{\bar\Omega}p(\mathbf x)}{\min_{\bar\Omega} p(\mathbf x)}\leq C_2(\gamma).
  \end{array}\ee
\end{theorem}
\textbf{Proof.} 
Inequalities (\ref{bouneq})$_1$, (\ref{bouneq})$_2$ follow easily from (\ref{bound}). In view of the Harnack inequality, (\ref{bouneq})$_3$, (\ref{bouneq})$_4$ are obtained.\\
\noindent
Let us set
$\alpha_1$ the lowest positive eigenvalue of the spectral problem
 \be\label{sp}
 \begin{cases}
  \Delta\varphi=-\alpha \varphi,\quad \mbox{in }\Omega\\
  \nabla\varphi\cdot\mathbf n=0,\quad \mbox{on }\partial\Omega\times\mathbb R^+
 \end{cases}\ee
 and 
 \be
 \bar\varphi=\dis\frac{1}{\vert\Omega\vert}\dis\int_\Omega\varphi\,d\Omega,\qquad \forall\varphi\in\{n,p\}.\ee
The following theorem provides a sufficient condition for the non-existence of non-constant solutions of (\ref{equi}).
\begin{theorem}
 If
 \be\label{nonexist}
\begin{cases}
 \gamma_1\geq \dis\frac{C_1(\gamma)}{2\alpha_1}\left\{\mu \rho +(1-\eta)\Phi+\sigma (1-\eta)^2\Phi \left[ M_1 +M_2\right]\right\},\\
 \gamma_2\geq \dis\frac{C_2(\gamma)\theta(1-\eta)}{4\mu^2\alpha_1}\left\{\mu \rho +(1-\eta)\Phi+\sigma (1-\eta)^2\Phi  M_1 +1\right\},
\end{cases}
 \ee
holds, then system (\ref{equi}) does not admit any positive non constant solution.
\end{theorem}
\textbf{Proof.} Let $(n,p)$ be a positive solution of (\ref{equi}). Multiplying (\ref{equi})$_1$ by $\dis\frac{n-\bar n}{n}$, (\ref{equi})$_2$ by $\beta \dis\frac{p-\bar p}{p}$, integrating over $\Omega$ and adding the resulting equations, one obtains -- in view of the divergence theorem and the boundary conditions (\ref{ibc})$_2$
\be\label{uno}
\begin{array}{l}
 \gamma_1\dis\int_\Omega \bar n \dis\frac{(\nabla n)^2}{n^2}d\Omega+\beta\gamma_2\dis\int_\Omega \bar p\dis\frac{(\nabla p)^2}{p^2}d\Omega=
 -\mu \rho\dis\int_\Omega \dis\frac{(n-\bar n)(p-\bar p)}{(1+\rho p)(1+\rho\bar p)}d\Omega\\
 -\|n-\bar n\|^2-(1-\eta)\Phi\int_\Omega
 \dis\frac{(n-\bar n)(p-\bar p)}{\left[1+\sigma (1-\eta)n+\xi p\right]\left[1+\sigma(1-\eta)\bar n+\xi\bar p\right]}d\Omega\\
 -\sigma(1-\eta)^2\Phi \bar n\dis\int_\Omega\dis\frac{(n-\bar n)(p-\bar p)}{\left[1+\sigma (1-\eta)n+\xi p\right]\left[1+\sigma(1-\eta)\bar n+\xi\bar p\right]}d\Omega\\
 +\sigma(1-\eta)^2\Phi\bar p\dis\int_\Omega \dis\frac{(n-\bar n)^2}{\left[1+\sigma (1-\eta)n+\xi p\right]\left[1+\sigma(1-\eta)\bar n +\bar p\right]}d\Omega\\
 +\theta\beta(1-\eta)\dis\int_\Omega \dis\frac{(n-\bar n)(p-\bar p)}{\left[(1-\eta)n+\nu\right]\left[(1-\eta)\bar n+\nu\right]}d\Omega.
\end{array}
\ee
Applying the Poincar\'e inequality and (\ref{bouneq})$_3$-(\ref{bouneq})$_4$, one recovers that
\be\label{due}
\begin{array}{l}
\!\!\!\! \gamma_1\dis\int_\Omega \bar n \dis\frac{(\nabla n)^2}{n^2}d\Omega\!+\!\beta\gamma_2\dis\int_\Omega \bar p\dis\frac{(\nabla p)^2}{p^2}d\Omega
 \!\geq\! \dis\frac{\alpha_1\gamma_1}{C_1(\gamma)}\|n\!-\!\bar n\|^2\!+\!\dis\frac{\alpha_1\beta\gamma_2}{C_2(\gamma)}\|p\!-\!\bar p\|^2.
\end{array}\ee
Subsituting (\ref{due}) in (\ref{uno}), in view of (\ref{bouneq})$_1$-(\ref{bouneq})$_2$, choosing $\beta=\dis\frac{2\nu^2}{\theta(1-\eta)}$,
one has that
\be\begin{array}{l}
  \dis\frac{\alpha_1\gamma_1}{C_1(\gamma)}\|n-\bar n\|^2+\dis\frac{\alpha_1\beta\gamma_2}{C_2(\gamma)}\|p-\bar p\|^2\leq\\
  \leq \left[\mu\rho+(1-\eta)\Phi+\sigma(1-\eta)^2\Phi M_1+2\sigma(1-\eta)^2\Phi M_2\right]\dis\frac{\|n-\bar n\|^2}{2}\\+
  \left[\mu\rho+(1-\eta)\Phi+\sigma(1-\eta)^2\Phi M_1+2\right]\dis\frac{\|p-\bar p\|^2}{2}
   \end{array}\ee
that is impossible when (\ref{nonexist}) holds.\\
\noindent
In the sequel we assume that (\ref{nonexist}) holds and hence (\ref{modadi}) admits only the constant steady states found in \cite{F}, i.e.:
\begin{itemize}
 \item $E_0=(0,0)$, representing the extinction of both species;
 \item $E_1=(\mu-\delta,0)$, the prey-only equilibrium;
 \item $E_2=(0,\nu/\theta)$, the predator-only equilibrium;
 \item $E_*=(n_*,p_*)$, the coexistence equilibrium, with $p_*=\dis\frac{(1-\eta)n_*+\nu}{\theta}$ and $n_*$ positive solution of
 \be
 U_1 n_*^3+U_2 n_*^2 + U_3 n_* +U_4=0,\ee
 where 
 \be\begin{array}{l}
     U_1=\rho (1-\eta)^2(\xi+\sigma\theta),\\
     U_2=(1-\eta)\rho \nu(2\xi+\sigma\theta)+ (1-\eta)\left\{(1-\eta)\left[\delta\rho\theta+\delta\rho\xi+
     \Phi\rho(1-\eta)\right]+\right.\\
     \left.+\sigma \theta^2+\rho\theta +\xi\theta\right\}>0,\\
     U_3=\nu^2\rho\xi+\nu\left\{(1-\eta)\left[\delta\rho\sigma\theta+2\delta\rho\xi+2\Phi\rho(1-\eta)\right]+\theta\rho+\theta\xi\right\}+\\
     +\theta\left\{(1-\eta)\left[(\xi+\sigma\theta)(\delta-\mu)+\delta\rho\right]+\Phi(1-\eta)^2+\theta\right\},\\
     U_4=\nu^2\rho \left[\delta\xi+\Phi(1-\eta)\right]+\theta \nu \left[\xi(\delta-\mu)+\Phi(1-\eta)+\delta\rho\right]+\theta^2(\delta-\mu).
    \end{array}\ee
\end{itemize}
We remark that, by the Descartes rules, if $U_4<0$, there exists at least one coexistence equilibrium. In particular:
\begin{itemize}
 \item if $\{U_3>0, U_4>0\}$, (\ref{modadi}) does not admit any coexistence equilibrium;
 \item if $\{U_3>0,U_4<0\}$ or $\{U_3<0,U_4<0\}$, (\ref{modadi}) admits a unique coexistence equilibrium;
 \item if $\{U_3<0,U_4>0\}$, (\ref{modadi}) admits two coexistence equilibria.
\end{itemize}
\section{Linear instability}
In this Section, we investigate the linear instability of the coexistence equilibrium. In particular, we look for conditions guaranteeing the stability in the absence of diffusion and instability driven by the diffusion (Turing instability). In this analysis we show that the sign of $a_{11}$ plays a fundamental role. In fact, $a_{11}>0$ is a necessary condition for the occurrence of such a kind of instability.
\subsection{Linear instability in the absence of diffusion}
The Jacobian matrix -- evaluated in $E_*$ -- is
\be\label{l0}
\mathcal L^0=\left(\begin{array}{cc}
           a_{11}&a_{12}\\
           a_{21}&-1
          \end{array}\right)
          \ee
          with 
\be\label{coeff}
\begin{cases}
 a_{11}=-n_*+\dis\frac{(1-\eta)^2\sigma\Phi n_*p_*}{\left[1+\sigma(1-\eta)n_*+\xi p_*\right]^2},\,\,\, a_{21}=\dis\frac{\theta (1-\eta)p_*^2}{\left[(1-\eta)n_*+\nu\right]^2}>0\\
a_{12}=-\dis\frac{\mu \rho n_*}{(1+\rho n_*)^2}-\dis\frac{\left[1+\sigma(1-\eta)n_*\right](1-\eta)\Phi n_*}{\left[1+\sigma(1-\eta)n_*+\xi p_*\right]^2}<0. 
\end{cases}\ee
In \cite{F} it has been proved that $a_{11}<0$ implies the linear stability of $E_*$. 
However, this is only a sufficient condition for the linear stability. 
In fact, setting
\be\label{invaode}
I_1^0=\mbox{tr} \mathcal L^0=a_{11}-1,\quad I_2^0=\mbox{det }\mathcal L^0=-a_{11}-a_{12}a_{21}\ee
the characteristic equation whose solutions are the $\mathcal L^0-$eigenvalues, is
\be\label{carode}
\lambda^2-I_1^0\lambda+I_2^0=0.\ee
Hence
 \be\label{staode2}
 a_{11}<\dis\min\{1,-a_{12}a_{21}\},\ee
 guarantees that $\{I_1^0<0,\, I_2^0>0\}$, i.e. the validity of the Routh-Hurwitz conditions  necessary and sufficient to guarantee that all the roots of (\ref{carode}) have negative real part  (\cite{mer}).
\subsection{Linear instability of E* in the presence of diffusion}
Setting 
\be
U_1=n-n^*,\qquad U_2=p-p^*,\ee
the linear system governing the evolution of perturbation fields to $E_*$,  
 is 
\be\label{5.7**}
\dis\frac{\partial \mathbf U}{\partial t}=\mathcal L^0\mathbf U+\mathcal D\mathbf U,\ee
where $\mathbf U=(U_1,U_2)^T$, $\mathcal L^0$ is given by (\ref{l0}) and $\mathcal D=\left(\begin{array}{cc}
                                                                   \gamma_1&0\\
                                                                   0&\gamma_2
                                                                  \end{array}\right)
$. The dispersion relation governing the eigenvalues $\lambda$ in terms of the wave number $k$ is 
\be\label{eqcar}
\lambda^2-T_k\lambda+h(k^2)=0,\ee
where
\be\label{usc}\begin{cases}
T_k=\mbox{tr}(\mathcal L^0)-k^2\mbox{tr}\mathcal D=\texttt I^0_1-k^2(\gamma_1+\gamma_2),\\
h(k^2)=\det \mathcal D k^4+k^2(\gamma_1-a_{11}\gamma_2)+\det \mathcal L^0=
\gamma_1\gamma_2 k^4+k^2(\gamma_1-a_{11}\gamma_2)+\texttt I^0_2.
\end{cases}
\ee
We remark that, if 
either
 \be\label{stabpde2}
a_{11}<\dis\min\left\{1,-a_{12}a_{21},\dis\frac{\gamma_1}{\gamma_2}\right\},\ee
or 
\be\label{stabpde3}
\begin{cases}
 0<a_{11}<\dis\min\left\{1,-a_{12}a_{21}\right\},\,\,\,\gamma_1<a_{11}\gamma_2,\\
 (\gamma_1-a_{11}\gamma_2)^2-4\gamma_1\gamma_2\texttt I^0_2<0,
\end{cases}
\ee
then $T_k<0,\,h(k^2)>0,\forall k$, i.e.
 $E_*$ -- stable in the absence of diffusion -- continues to be stable in the presence of diffusion too.
\\
\noindent
From \eqref{staode2} and \eqref{stabpde2}, the condition $a_{11}<0$ implies stability in the absence and in the presence of diffusion. Hence, if we are looking for conditions guaranteeing the diffusion-driven instability, we have to explore the dynamics in the case $0<a_{11}<\min\{1,-a_{12}a_{21}\}$.
Since $\texttt I^0_1<0\Rightarrow T_k<0,\forall k$, for the occurrence of Turing instability, it is sufficient that $h(k^2)$ assumes some negative value (i.e. its minimum is negative).
In view of 
\be
\dis\frac{\partial h(k^2)}{\partial k^2}=2\gamma_1\gamma_2k^2+\gamma_1-a_{11}\gamma_2,\ee
it turns out that the minimum of $h(k^2)$ is obtained for
\be\label{alphamin}
(k^2)_{\mbox{min}}=\dis\frac{a_{11}\gamma_2-\gamma_1}{2\gamma_1\gamma_2}.\ee
From the positive definiteness of $k^2$ it follows that, a \emph{necessary} condition for the occurrence of Turing instability is 
\be\label{cnectur}
\dis\frac{\gamma_1}{\gamma_2}<a_{11}.\ee
Obviously, \eqref{cnectur} requires that $a_{11}>0$ in order to be satisfied. \\
The minimum of $h(k^2)$ is 
\be\label{mini2}
(h(k^2))_{\mbox{min}}=\texttt I^0_2-\dis\frac{(a_{11}\gamma_2-\gamma_1)^2}{4\gamma_1\gamma_2}.\ee
Hence $h(k^2)$ assumes some negative value if $ \texttt I^0_2<\dis\frac{(a_{11}\gamma_2-\gamma_1)^2}{4\gamma_1\gamma_2}$. Summarizing, 
 \be\label{tur}
 \begin{cases}
  0<a_{11}<\dis\min\{1,-a_{12}a_{21}\},\quad \gamma_1<a_{11}\gamma_2,\\
  \texttt I^0_2<\dis\frac{(a_{11}\gamma_2-\gamma_1)^2}{4\gamma_1\gamma_2},
 \end{cases}\ee
 guarantees that Turing instability occurs.

To the bifurcation, $(\texttt I_{2i})_{\mbox{min}}=0$. Setting $\gamma=\dis\frac{\gamma_1}{\gamma_2}$, it turns out that the critical value of $\gamma$ at the bifurcation, is 
\be\label{gammacritico}
\gamma_c=(2\texttt I^0_2+a_{11})^2-2\sqrt{\texttt I^0_2(\texttt I^0_2+a_{11})}
\ee
and the critical wave number is
\be
k^2_c=\dis\frac{a_{11}-\gamma_c}{2\gamma_1}.\ee
For $\gamma>\gamma_c$, the range of the wave number for the instability, is 
\be
k^2_{-}<k^2<k^2_+,\ee
with
\be\begin{cases}
k^2_{-}=\dis\frac{-(\gamma_1-a_{11}\gamma_2)-\sqrt{\Delta}}{2\gamma_1\gamma_2},
\,\,
k^2_+=\dis\frac{-(\gamma_1-a_{11}\gamma_2)+\sqrt{\Delta}}{2\gamma_1\gamma_2},\\
\Delta=(\gamma_1-a_{11}\gamma_2)^2-4\gamma_1\gamma_2\texttt I^0_2.\end{cases}\ee
Investigation shows that the set of biologically meaningful parameters verifying \eqref{tur} is not empty but small. Then, in order to better explore the pattern formation in the most significant case $a_{11}<0$, model \eqref{mod} needs to be generalized. To this aim, in the following section, we investigate the influence of linear cross-diffusion terms on the population dynamics.
\section{
Cross-diffusion driven instability }
When both linear self and cross-diffusion terms are introduced,
the linear system  (\ref{5.7**}) can be rewritten as follows
\be\label{5.7***}
\dis\frac{\partial \mathbf U}{\partial t}=\mathcal L^0\mathbf U+ \mathcal D^\prime \Delta \mathbf U ,\ee
where
\be\label{elle0}\mathcal L^0=
\left(\begin{array}{cc}
                                                 a_{11}&a_{12}\\
                                                 a_{21}&-1
                    \end{array}
\right), \,\,\,\,\mathcal D^\prime=
\left(\begin{array}{cc}
                                                 \gamma_{11}\,&\gamma_{12}\\
                                                 \gamma_{21}\,&\gamma_{22}
                    \end{array}
\right), \ee
$\gamma_{11}=\gamma_1,\gamma_{22}=\gamma_2$ and $\texttt det \mathcal D^\prime >0$.
The dispersion relation (\ref{eqcar}) which gives the eigenvalue $\lambda$ in terms of the wave number $k$ is 
\be\label{eqcarc}
\lambda^2-\texttt T^\prime_{k}\lambda+\texttt h^\prime(k^2)=0,\ee
where
\be\label{inv}\begin{cases}
\texttt T^\prime_{k}=\texttt tr({\mathcal L^0}) -k^2\texttt tr(\mathcal D^\prime)=\texttt I^0_1-k^2(\gamma_{11}+\gamma_{22})=\texttt T_k,\\
\texttt h^\prime(k^2)=\texttt det \mathcal D^\prime k^4+q^\prime k^2+\texttt det {\mathcal L^0}
=(\gamma_{11}\gamma_{22}-\gamma_{12}\gamma_{21})k^4+q^\prime k^2+\texttt I^0_2,
\\
q^\prime=\gamma_{11}-a_{11}\gamma_{22}+a_{12}\gamma_{21}+a_{21}\gamma_{12}.
\end{cases}
\ee
We are looking for  those modes $k\neq 0$  such that $ h^\prime(k^2)<0.$
The only possibility for $ h^\prime(k^2)<0$ is
requiring $q^\prime < 0$. 
The condition for the marginal stability at some $k^2=k^2_{cr}$ is $\min(h^\prime(k^2_{cr}))=0$ and the minimum of $h^\prime$ is reached at $k^2_{cr}=-\frac{q^\prime}{2 det D^\prime}.$
In addition $h^\prime(k^2_{cr})<0$ gives 
${q^\prime}^2-4 \texttt I_2^0 \texttt det D^\prime>0.$

The conditions for cross-diffusion-driven instability of system (\ref{5.7***}),(\ref{ibc}) around the homogeneous
steady state $ E_*$ can be summarized as follows
\be\label{60}
\begin{cases}
a_{11}-1<0, \qquad  -a_{11}-a_{12}a_{21}>0,\\
a_{21} \gamma_{12}+a_{12} \gamma_{21}+\gamma_{11}-a_{11} \gamma_{22}<0,\\
\gamma_{11}\gamma_{22}-\gamma_{12}\gamma_{21}>0,\\
(a_{21} \gamma_{12}+a_{12} \gamma_{21}+\gamma_{11}-a_{11} \gamma_{22})^2+4 (a_{11}+a_{12}a_{21})(\gamma_{11}\gamma_{22}-\gamma_{12}\gamma_{21})>0.
\end{cases}
\ee
The above inequalities \eqref{60} define a region where the coexistence equilibrium E* is unstable. 
Choosing $\gamma_{12}$ as bifurcation parameter and  $\gamma_{12}= \gamma_{12}^{cr}$ as Turing threshold, bifurcation
happens at the critical value  
\be \label{gammacr}
\gamma_{12}^{cr}= \dis \frac{A+\sqrt{B}}{a_{21}^2}
\ee
where
\be
A=a_{21}(a_{12} \gamma_{21}+a_{11} \gamma_{22}- \gamma_{11})+2 a_{11}\gamma_{21}
\ee
\be
\begin{array}{l}
B=2a^2_{21}(-2 \gamma_{11}a_{12} \gamma_{21}+a_{12} a_{11}\gamma_{21}]\gamma_{22}-2 a_{11}\gamma_{11}\gamma_{22}-2a_{12} a_{21}\gamma_{11}\gamma_{22})+4 a^2_{11}\gamma^2_{21}\\
- 4a_{21}( a_{11}\gamma_{11}\gamma_{21}-a_{12}a_{11}\gamma^2_{21}-a^2_{11}\gamma_{22}\gamma_{21})
\end{array}
\ee
corresponding with the critical wavenumber
\be
k^2_{cr}=\dis \sqrt{ -\frac{a_{11}+a_{12}a_{21}}{\gamma_{11}\gamma_{22}-\gamma_{12}\gamma_{21}}} \neq 0
\ee
For $\gamma_{12}>\gamma_{12}^{cr}$ the unstable
wavenumbers stay in between the roots $k^2_{-}, \, k^2_{+}$ roots of $h^\prime(k^2)=0.$

\section{Amplitude equations and stability of spatial patterns}
To obtain the intervals of control parameters for different kinds of spatial patterns - which provide information on  inhomogeneous distribution of both populations on
the whole domain - we need to derive and analyze via multiple scale analysis the amplitude equations. The well-known amplitude equations are obtained via the standard method. Here we give the main steps.
We consider the following system and take $\gamma_{12}$ as a Turing bifurcation parameter
 \[   
  \dis\left(\begin{array}{l}
                              \dis       \frac{\partial n}{\partial t} \\
\\
                               \dis     \frac{\partial p}{\partial t}  
                                     \end{array}
\right)=L(\gamma_{12})\left(\begin{array}{l}
                         n\\
                                      p
                                                             \end{array}
\right)+\frac12\left(\begin{array}{l}
               f_{nn}n^2+2f_{np}np+f_{pp}p^2\\
               g_{nn}n^2+2g_{np}np+g_{pp}p^2          \end{array}
\right)\,\,\,\,\,\,\,\,\,\,\,\,\,\,\,\,\,\,\,\,\,\,\,\,\,\,\,\,\,\,\,\,\,\,\,\,\,\,\,\,\,\,\,\,\,\,\,\,\,\,\,\,\,\,\,\,\,\,\,\,\,\,\,\,\,
\]
\be \label{wna}\,\,\,\,\,\,\,\,\,\,\,\,\,\,\,\,\,\,\,\,\,\,\,\,\,\,\,\,\,\,\,\,\,\,\,\,\,\,\,\,\,\,\,\,\,\,\,\,\,\,\,\,+\frac16\left(\begin{array}{l}
               f_{nnn}n^3+3f_{nnp}n^2p+f_{npp}np^2+f_{ppp}p^3\\
               g_{nnn}n^3+3g_{nnp}n^2p+g_{npp}np^2+g_{ppp}p^3          \end{array}
\right),\ee
where we take the linear operator 
\be
L(\gamma_{12})=\left(\begin{array}{cc}
                   a_{11}+\gamma_{11}\Delta \,\,\,&a_{12}+\gamma_{12}\Delta\\
                   a_{21}+\gamma_{21}\Delta \,\,\,&a_{22}+\gamma_{22}\Delta
                  \end{array}
\right)
\ee
the expression of $a_{ij}$ are given in (\ref{coeff}) 
and
\be
\begin{array}{l}
f_{nn}=-2-\dis \frac{-2 n^2(1+a n^{\alpha})^3 +n^{\alpha} p \alpha (1-\alpha + a n^{\alpha}(1+\alpha))}{n^2(1+a n^{\alpha})^3}, \,\,  f_{pp}=0,\\
f_{np}=\dsp-\frac{n^{-1+\alpha}\alpha}{(1+a n^{\alpha})^2}, \,\,\,\,\, g_{pp}=g_{ppp}=g_{npp}=0,\,\,\,\,
 g_{np}=\dsp \frac{n^{-1+\alpha}\alpha \gamma }{(1+a n^{\alpha})^2},\\
 g_{nn}=\dsp-\frac{n^{-2+\alpha}\alpha p(1-\alpha + a n^{\alpha}(1+\alpha))\gamma}{(1+a n^{\alpha})^3}, \,\,\,\,
 g_{nnn}=\dsp \frac{1}{(1+a n^{\alpha})^4},\\ 
  g_{nnp}=\dsp \frac{n^{-2+\alpha}\alpha (1-\alpha + a n^{\alpha}(1+\alpha))}{(1+a n^{\alpha})^3},\,\,f_{nnn}=\dsp-\frac{1}{{(1+a n^{\alpha})^4}},\\ 
  f_{ppp}=f_{npp}=0,\,\,\,\,\,
 f_{nnp}=\dsp \frac{n^{-2+\alpha}\alpha (1-\alpha + a n^{\alpha}(1+\alpha))}{(1+a n^{\alpha})^3}.
\end{array}
\ee
At the onset of Turing instability, the solution of our problem can be expanded 
\be
 \mathbf X= \mathbf X_s+\dsp \sum_{j=1}^3  \mathbf X_0[A_j\exp(i\mathbf  k_j\cdot\mathbf r)+\bar A_j \exp(-i\mathbf  k_j\cdot\mathbf r)]
\ee
where $\mathbf X_s$ represents the uniform steady state, $\mathbf X_0$ the direction of eigenmodes and $A_j, \bar A_j$ the amplitudes associated with the modes $\mathbf  k_j$, $-\mathbf  k_j.$
Introducing the additional small parameter $\epsilon$, near the Turing critical value  $\gamma_{12}^{cr}$ , 
we perturb the bifurcation parameter $\gamma_{12}$ along $n$, $p$, $t$ \be
\begin{array}{l}
\gamma_{12}=\gamma_{12}^{cr}+\epsilon \gamma_{12}^{(1)}+\epsilon^2 \gamma_{12}^{(2)}+\epsilon^3 \gamma_{12}^{(3)}+...\\
n=\epsilon n_1+\epsilon^2 n_2+\epsilon^3n_3+...\\
p=\epsilon p_1+\epsilon^2 p_2+\epsilon^3p_3+...\\
t=t_0+\epsilon t_1+\epsilon^2 t_2+...
\end{array}\ee
This leads to
\be
L(\gamma_{12}(\epsilon))=L^{cr}+\epsilon \gamma_{12}^{(1)} \left(\begin{array}{ll}
                         0& \Delta\\
                                     0 &0
                                                             \end{array}
\right)+\epsilon^2 \gamma_{12}^{(2)} \left(\begin{array}{ll}
                         0& \Delta\\
                                     0 &0
                                                             \end{array}
\right) +o(\epsilon^2),
\ee
where
\be   \label{cri}
L^{cr}=\left(\begin{array}{cc}
                   a_{11}+\gamma_{11}\Delta\,\,\,\,\,\,\,&a_{12}+\gamma_{12}^{cr}\Delta\\
                   a_{21}+\gamma_{21}\Delta\,\,\,\,\,\,\,&a_{22}+\gamma_{22}\Delta
                  \end{array}
\right),
\ee
To apply the multiple scale method we introduce
$ t_0=t,\, t_1=\epsilon t,\,\,t_2=\epsilon^2 t,$ and we obtain 
\be
\dsp \frac{\partial}{\partial t}= \frac{\partial}{\partial t_0}+ \epsilon \frac{\partial}{\partial t_1}+\epsilon^2 \frac{\partial}{\partial t_2}+o(\epsilon^2)
\ee
From (\ref{wna}) and balancing the coefficients of $\epsilon^j$, we have\\
at first order
\be \label{uno1}
L^{cr}\left( \begin{array}{l}
               n_1\\
               p_1
              \end{array} \right)=\left( \begin{array}{l}
               0\\
               0
              \end{array} \right)
              \ee
at second order
\[
L^{cr}\left( \begin{array}{l}
               n_2\\
               p_2
              \end{array} \right)=\dsp \frac{\partial}{\partial t_1}\left( \begin{array}{l}
               n_1\\
               p_1
              \end{array} \right)-  \left(\begin{array}{ll}
                         0& \gamma_{12}^{(1)}\\
                                     0 &\,\,0
                                                             \end{array}
\right)+\Delta \left( \begin{array}{l}
              n_1\\
                 p_1
              \end{array} \right)
              \]
\be  \label{due2}
-\frac12\left(\begin{array}{l}
               f_{nn}n_1^2+2f_{np}n_1p_1+f_{pp}p_1^2\\
               g_{nn}n_1^2+2g_{np}n_1p_1+g_{pp}p_1^2          \end{array}
\right) =\left( \begin{array}{l}
               F_n\\
               F_p
              \end{array} \right)
\ee
at third order
\[
L^{cr}\left( \begin{array}{l}
               n_3\\
               p_3
              \end{array} \right)=\left( \begin{array}{ll}
                \frac{\partial n_2}{\partial t_1}+ \frac{\partial n_1}{\partial t_2}\\
                \frac{\partial p_2}{\partial t_1}+ \frac{\partial p_1}{\partial t_2}\
              \end{array} \right)-   \left(\begin{array}{ll}
                         0&\, \gamma_{12}^{(1)}\\
                                     0 &\,\,\,0
                                                             \end{array}
\right)\Delta\left( \begin{array}{l}
              n_2\\
                 p_2
              \end{array} \right)-   \left(\begin{array}{ll}
                         0& \gamma_{12}^{(1)}\\
                                     0 &\,\,0
                                                             \end{array}
\right)\Delta \left( \begin{array}{l}
              n_1\\
                 p_1
              \end{array} \right)
              \]
              
              \be  \label{tre}
              -\left(\begin{array}{l}
               f_{nn}n_1n_2+f_{np}(n_1p_2+n_2p_1)+f_{pp}p_1p_2\\
                g_{nn}n_1n_2+g_{np}(n_1p_2+n_2p_1)+g_{pp}p_1p_2        \end{array}
\right)\ee
\[ - \frac16\left(\begin{array}{l}
               f_{nnn}n_1^3+3f_{nnp}n_1^2p_1+3f_{npp}n_1p_1^2+f_{ppp}p_1^3\\
               g_{nnn}n_1^3+3g_{nnp}n_1^2p_1+3g_{npp}n_1p_1^2+g_{ppp}p_1^3       \end{array}
\right) =\left( \begin{array}{l}
               G_n\\
               G_p
              \end{array} \right)
\]
Solving (\ref{uno1}) we obtain 
\be
\left( \begin{array}{l}
               n_1\\
               p_1
              \end{array} \right)=\left( \begin{array}{l}
               \phi\\
               1
              \end{array} \right)\left(  \sum_{j=1}^3 W_j\exp(i\mathbf  k_j\cdot\mathbf r)+c.c.\right),
\ee
where c.c. denotes the complex conjugate of the previous terms, $W_j$ is the amplitude of the mode $\exp(i\mathbf  k_j\cdot\mathbf r)$ (j=1,2,3) and  $\phi=\dsp \frac{a_{12}-\gamma_{12}^{cr} k^2_{cr}}{\gamma_{11}k^2_{cr}-a_{11} }$.

According to the Fredholm solvability condition, the functions of the right-hand side of (\ref{due2}) must be orthogonal to the eigenvectors of the zero eigenvalue of  $\bar L_T$ which is the adjoint operator of $L_T.$ The eigenvectors of the operator $\bar L_T$ are $\left( \begin{array}{l}
              1\\
               \psi
              \end{array} \right) \exp(-i\mathbf  k_j\cdot\mathbf r)+c.c. (j=1,2,3)$ with $\psi=\dsp \frac{a_{12}-\gamma_{12}^{cr}k^2_{cr}}{\gamma_{22}k^2_{cr}+1}.$ 

The orthogonality condition  is
              \[
              (1, \psi)\left( \begin{array}{l}
              F_n^j\\
               F_p^j
              \end{array} \right)=0, \,\,\,(j=1,2,3)
               \]
where $F_n^j$ and $F_p^j$  give the coefficients of $ \exp(i\mathbf  k_j\cdot\mathbf r)$ in $F_n$ and $F_p$. From this relation it follows 
\be
\label{orteq}
\begin{cases}
(\phi+\psi) \dsp\frac{ \partial W_1}{\partial t_1}=-k^2_{cr}\gamma_{12}^{(1)} W_1+(f_2+\psi g_2) \bar W_2 \bar W_3, \\
(\phi+\psi) \dsp\frac{ \partial W_2}{\partial t_1}=-k^2_{cr}\gamma_{12}^{(1)} W_2+(f_2+\psi g_2) \bar W_3 \bar W_1,\\
(\phi+\psi) \dsp\frac{ \partial W_3}{\partial t_1}=-k^2_{cr}\gamma_{12}^{(1)} W_3+(f_2+\psi g_2) \bar W_1 \bar W_2, \end{cases}
\ee
where
\be
\label{orteq2}
\begin{cases}
f_2=f_{nn}\phi^2+2 f_{np}\phi+f_{pp},\\
g_2=g_{nn}\phi^2+2 g_{np}\phi+g_{pp}.
  \end{cases}
\ee
Following a similar procedure for (\ref{due2}) 
 its solution will be  of type
\[
\left( \begin{array}{l}
               n_2\\
               p_2
              \end{array} \right)=\left( \begin{array}{l}
               N_0\\
               P_0
              \end{array} \right)+\sum_{j=1}^3 \left( \begin{array}{l}
               N_j\\
               P_j
              \end{array} \right)\exp(i\mathbf  k_j\cdot\mathbf r)+\sum_{j=1}^3 \left( \begin{array}{l}
               N_{jj}\\
               P_{jj}
              \end{array} \right)\exp(2i\mathbf  k_j\cdot\mathbf r)
              \]
             \[
              +\sum_{j=1}^3 \left( \begin{array}{l}
               N_{12}\\
               P_{12}
              \end{array} \right)\exp(i(\mathbf  k_1-\mathbf  k_2)\cdot\mathbf r)+\sum_{j=1}^3 \left( \begin{array}{l}
               N_{23}\\
               P_{23}
              \end{array} \right)\exp(i(\mathbf  k_2-\mathbf  k_3)\cdot\mathbf r)
              \]
              \be
              +\sum_{j=1}^3 \left( \begin{array}{l}
               N_{31}\\
               P_{31}
              \end{array} \right)\exp(i(\mathbf  k_3-\mathbf  k_1)\cdot\mathbf r)+c.c.
\ee
Substituting in (\ref{due2}), separating the coefficients of $\exp(0), \, \exp(i\mathbf  k_j\cdot\mathbf r), \, \exp(2i\mathbf  k_j\cdot\mathbf r), \,\exp(i(\mathbf  k_1-\mathbf  k_2)\cdot\mathbf r)$ (and permuting the suffixes we obtain also the coefficients corresponding to $\exp(i(\mathbf  k_2-\mathbf  k_3)\cdot\mathbf r), \,\,\,\exp(i(\mathbf  k_3-\mathbf  k_1)\cdot\mathbf r)$) denoting by
\be
{\cal M}_{k}=\left(\begin{array}{ll}
               a_{11}-\gamma_{11}k^2 \,\,\,\, &\,\,a_{12}-\gamma_{12}^{cr}k^2\\
               a_{21}-\gamma_{21}k^2\,\,\,\,  &\,\,-1-\gamma_{22}k^2     \end{array}
\right)
\ee
we get, for $N_j=\phi P_j, \,\,\,j=1,2,3$
\[
\left( \begin{array}{l}
               N_0\\
               P_0
              \end{array} \right)={\cal M}_{0}^{-1}
           \left( \begin{array}{l}
              - f_2\\
               -g_2
              \end{array} \right)(\vert W_1 \vert^2+\vert W_2 \vert^2+\vert W_3 \vert^2)= \left( \begin{array}{l}
               Z_{n0}\\
               Z_{p0}
              \end{array} \right)(\vert W_1 \vert^2+\vert W_2 \vert^2+\vert W_3 \vert^2),
              \]
\[
\left( \begin{array}{l}
               N_{11}\\
               P_{11}
              \end{array} \right)={\cal M}_{2k_{cr}}^{-1}
           \left( \begin{array}{l}
              - \frac{f_2}{2}\\
              -  \frac{g_2}{2}
              \end{array} \right)W_1^2= \left( \begin{array}{l}
               Z_{n1}\\
               Z_{p1}
              \end{array} \right)W_1^2,
              \]
              \[
              \left( \begin{array}{l}
               N_{12}\\
               P_{12}
              \end{array} \right)={\cal M}_{\sqrt 3k_{cr}}^{-1}
           \left( \begin{array}{l}
              - f_2\\
              -  g_2
              \end{array} \right)W_1\bar W_2= \left( \begin{array}{l}
               Z_{n2}\\
               Z_{p2}
              \end{array} \right)W_1\bar W_2.
              \]
At third order,  collecting the coefficients $(G_n^1,G_p^1)^T$ of $ \exp(i\mathbf  k_1\cdot\mathbf r)$ from (\ref{tre}), we find
\[
 \left( \begin{array}{l}
               G_n^1\\
               G_p^1
              \end{array} \right)=\left( \begin{array}{ll}
               \phi( \frac{\partial P_1}{\partial t_1}+ \frac{\partial W_1}{\partial t_2})\\
               \,\,\, \frac{\partial P_1}{\partial t_1}+ \frac{\partial W_1}{\partial t_2}\
              \end{array} \right)- \left(\begin{array}{ll}
                         0& \gamma_{12}^{(1)}\\
                                     0 &\,\,\,0
                                                             \end{array}
\right)\left( \begin{array}{l}
           \,\, \,\,  \phi P_1\\
               - k^2_{cr} P_1
              \end{array} \right)-\left(\begin{array}{ll}
                         0& \gamma_{12}^{(2)}\\
                                     0 &\,\,\,0
                                                             \end{array}
\right)\left( \begin{array}{l}
            \,\,\,  \phi W_1\\
              - k^2_{cr}   W_1
              \end{array} \right)
              \]
              \[ 
    \small \small { \!\!\!\!\!-\left(\begin{array}{l}
              [ (f_{nn}\phi +f_{np})(Z_{n0}+Z_{n1})+ (f_{np}\phi +f_{pp})(Z_{p0}+Z_{p1}) ]\vert W_1 \vert^2
               + [(f_{nn}\phi +f_{np})(Z_{n0}+Z_{n2})\\ 
              \qquad  \qquad +(f_{np}\phi +f_{pp})(Z_{p0}+Z_{p2})(\vert W_2 \vert^2+\vert W_3 \vert^2)]W_1
                +f_2(\bar W_2 \bar P_3+\bar W_3 \bar P_2)      \\
                  ( (g_{nn}\phi +g_{np})(Z_{n0}+Z_{n1})+ (g_{np}\phi +g_{pp})(Z_{p0}+Z_{p1}) )\vert W_1 \vert^2
               + [(g_{nn}\phi +g_{np})(Z_{n0}+Z_{n2})\\
              \qquad  \qquad +(g_{np}\phi +g_{pp})(Z_{p0}+Z_{p2})(\vert W_2 \vert^2+\vert W_3 \vert^2)]W_1
                +g_2(\bar W_2 \bar P_3+\bar W_3 \bar P_2)    \end{array}
           \right)}\]
\be - \left(\begin{array}{l}
           (\vert W_1 \vert^2+\vert W_2 \vert^2+\vert W_3 \vert^2  )  (f_{nnn}\phi^3+3f_{nnp}\phi^2+3f_{npp}\phi+f_{ppp}\\
                (\vert W_1 \vert^2+\vert W_2 \vert^2+\vert W_3 \vert^2  )  (g_{nnn}\phi^3+3g_{nnp}\phi^2+3g_{npp}\phi+g_{ppp}      \end{array}
\right)W_1
\ee
Analogously, permutating the subscript of $W$ and $P$, we can find the other coefficients $(G_n^2,G_p^2)^T,\,(G_n^3,G_p^3)^T.$
From Fredholm solvability condition $(1,\psi)\left( \begin{array}{l}
               G_n^j\\
                G_p^j
              \end{array} \right)=0,\quad j=1,2,3$
it follows that 
\be
\label{orteq3}
\begin{cases}
(\phi+\psi) \dsp(\frac{ \partial W_1}{\partial t_2}+\frac{ \partial P_1}{\partial t_1})=-k^2_{cr}(\gamma_{12}^{(1)}  P_1+\gamma_{12}^{(2)}  W_1)\\
\qquad \quad+h_1(\bar W_2 \bar P_3+ \bar W_3 \bar P_2)-(G_1 \vert W_1 \vert^2+G_2 (\vert W_2 \vert^2+\vert W_3 \vert^2  ) ) W_1\\
(\phi+\psi) \dsp(\frac{ \partial W_2}{\partial t_2}+\frac{ \partial P_2}{\partial t_1})=-k^2_{cr}(\gamma_{12}^{(1)} P_2+\gamma_{12}^{(2)} W_2)\\
\qquad\quad+h_1(\bar W_3 \bar P_1+ \bar W_1 \bar P_3)-(G_1 \vert W_2 \vert^2+G_2 (\vert W_3 \vert^2+\vert W_1 \vert^2  ) ) W_2\\
(\phi+\psi) \dsp(\frac{ \partial W_3}{\partial t_2}+\frac{ \partial P_3}{\partial t_1})=-k^2_{cr} (\gamma_{12}^{(1)} P_3+\gamma_{12}^{(2)} W_3)\\
\qquad \quad+h_1(\bar W_1 \bar P_2+ \bar W_2 \bar P_1)-(G_1 \vert W_3 \vert^2+G_2 (\vert W_1 \vert^2+\vert W_2 \vert^2  ) ) W_3\\
 \end{cases}
\ee
with
\be
\label{G}
\begin{array}{l}
h_1=f_2+\psi g_2\\
G_1=-[(f_{nn}\phi+f_{np}+\psi(g_{nn}\phi+g_{np}))(Z_{n0}+Z_{n1})\\\qquad+(f_{np}\phi+f_{pp}+\psi(g_{np}\phi+g_{pp}))(Z_{p0}+Z_{p1})+f_3+\psi g_3]
\\
G_2=-[(f_{nn}\phi+f_{np}+\psi(g_{nn}\phi+g_{np}))(Z_{n0}+Z_{n2})\\
\qquad +(f_{np}\phi+f_{pp}+\psi(g_{np}\phi+g_{pp}))(Z_{p0}+Z_{p2})+f_3+\psi g_3]
\\
f_3=f_{nnn}\phi^3+3f_{nnp}\phi^2+3f_{npp}\phi+f_{ppp}
\\
g_3=g_{nnn}\phi^3+3g_{nnp}\phi^2+3g_{npp}\phi+g_{ppp}
\end{array}
\ee
Denoting by  $A_j$ the amplitude and expanding as follows
\[
A_j=\epsilon W_j+\epsilon^2 V_j+o(\epsilon^2)
\]  from
\be \label{Aj}
\dsp \frac{\partial A_j}{\partial t}=\epsilon \frac{\partial A_j}{\partial t_1}+\epsilon^2\frac{\partial A_j}{\partial t_2}+o(\epsilon^2).
\ee
we obtain the amplitude equations 
\be
\label{ampeq}
\begin{cases}
\tau_0 \dsp\frac{ \partial A_1}{\partial t}=\mu A_1+h\bar A_2 \bar A_3- (b_1 \vert A_1 \vert^2+b_2(\vert A_2\vert^2+\vert A_3 \vert^2))A_1 \\
\tau_0 \dsp\frac{ \partial A_2}{\partial t}=\mu A_2+h\bar A_3 \bar A_1- (b_1 \vert A_2 \vert^2+b_2(\vert A_3\vert^2+\vert A_1 \vert^2))A_2 \\
\tau_0 \dsp\frac{ \partial A_3}{\partial t}=\mu A_3+h\bar A_1 \bar A_2- (b_1 \vert A_3 \vert^2+b_2(\vert A_1\vert^2+\vert A_2 \vert^2))A_3 
\end{cases}
\ee
where 
$$\tau_0=-\dsp \frac{(\phi+\psi)}{k^2_{cr}\gamma_{12}^{cr}},\,\,\,\mu=\dsp \frac{\gamma_{12}-\gamma_{12}^{cr}}{\gamma_{12}^{cr}},\,\,\, h=-\dsp \frac{h_1}{k^2_{cr} \gamma_{12}^{cr}},\,\,\,b_1=-\dsp \frac{G_1}{k^2_{cr} \gamma_{12}^{cr}},\,\,\,b_2=-\dsp \frac{G_2}{k^2_{cr}\gamma_{12}^{cr}}
$$
Each amplitude can be expressed through  a mode $\rho_j=\vert A_j\vert $ and a corresponding phase angle $\theta_j$ as $A_j=\rho_j \exp(i\theta_j), \,\,\,j=1,2,3.$  Substituting in (\ref{ampeq}) and separating the real and imaginary parts we obtain
\be
\label{ampl}
\begin{cases}
\tau_0 \dsp\frac{ \partial \theta}{\partial t}=-h\dsp\frac{\rho_1^2 \rho_2^2+\rho_1^2\rho_3^2+\rho_2^2 \rho_3^2}{\rho_1 \rho_2 \rho_3}\sin \theta\\
\tau_0 \dsp\frac{ \partial \rho_1}{\partial t}=\mu \rho_1+h \rho_2 \rho_3 \cos \theta- b_1 \rho_1^3-b_2(\rho_2^2+\rho_3^2) \rho_1\\
\tau_0 \dsp\frac{ \partial \rho_2}{\partial t}=\mu \rho_2+h \rho_3 \rho_1 \cos \theta- b_1 \rho_2^3-b_2(\rho_3^2+\rho_1^2) \rho_2\\
\tau_0 \dsp\frac{ \partial \rho_3}{\partial t}=\mu \rho_3+h \rho_1 \rho_2 \cos \theta- b_1 \rho_3^3-b_2(\rho_1^2+\rho_2^2) \rho_3
\end{cases}
\ee
with $\theta=\theta_1+\theta_2+\theta_3$. 
The above dynamical system admits the following different kinds of solutions:
\begin{itemize}
\item The homogeneous stationary state represented by 
\be
\rho_1=\rho_2=\rho_3=0
\ee
which is stable for $\mu<\mu_2=0$ and unstable for $\mu>\mu_2=0.$
\item Stripe pattern represented by
\be \label{stripe}
\rho_1=\sqrt{\frac{\mu}{b_1}}\neq 0, \qquad  \rho_2=\rho_3=0,
\ee
which are stable for $b_1<b_2$ and $\mu>\mu_3=\dsp\frac{b_1 h^2}{(b_2-b_1)^2}.$
\item Hexagonal pattern represented by 
\be
\rho_1=\rho_2=\rho_3=\rho_H^{\pm}=\dsp \frac{\vert h \vert \pm\sqrt{h^2+4(b_1+2b_2)\mu}}{2(b_1+2b_2)};
\ee
which exist and are stable when $2(b_1+2b_2)>0$  and $\mu>\mu_H=-\dsp \frac{h^2}{4(b_1+2b_2)}.$
 Therefore, the hexagons for $\rho_H=\rho_H^{+}$  are stable if $\mu<\mu_{H2}=\dsp\frac{(2b_1+b_2)h^2}{(b_2-b_1)^2},$ while for the solution $\rho_H=\rho_H^{-}$  the hexagonal structures are unstable.
\item Mixed state given by 
\be
\rho_1=\dsp \frac{\vert h \vert}{b_2-b_1}, \,\,\,\,\, \rho_2=\rho_3=\dsp \sqrt{\frac{\mu-b_1 \rho_1^2}{b_1+b_2}}
\ee
with $\mu>b_1 \rho_1^2$ and $b_2>b_1$ and is always unstable.
\end{itemize}

\begin{table}[]
    \centering
    \begin{tabular}{|c|ccccccc|}
     \hline
         Parameters & $\delta$ & $\phi$ & $\sigma$ & $\xi$& $\theta$ & $\nu$ & $\eta$\\
\hline
Values & 0.5& 2&2&3&0.95&0.1&0.2\\
\hline
    \end{tabular}
    \caption{Fixed values for some model parameters}
    \label{tab:par}
\end{table}

\section{Numerical Simulations}
\begin{figure}
\includegraphics[width=0.48\textwidth]{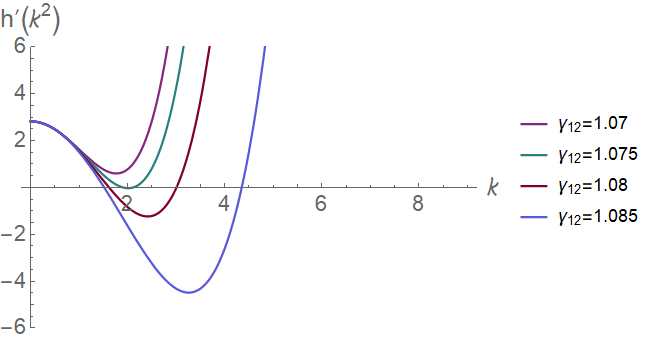}
\includegraphics[width=0.48\textwidth]{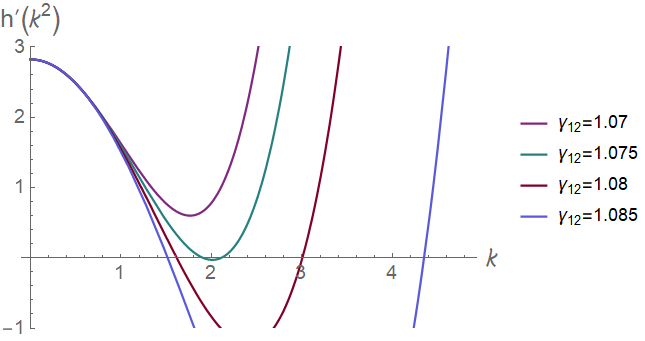}
\caption{\label{fig:ic}
Left panel: plots of $h'(k^2)$ as a function of the wavenumber $k$ for different values of the
bifurcation parameter $\gamma_{12}$; in the right panel, a detail of the same plots. Here $\gamma_{11} =3,\,\gamma_{21} =11,\,\gamma_{22} =4, $ and other parameter values as in Table \ref{tab:par} }
\end{figure}

\begin{figure}
\includegraphics[width=0.48\textwidth]{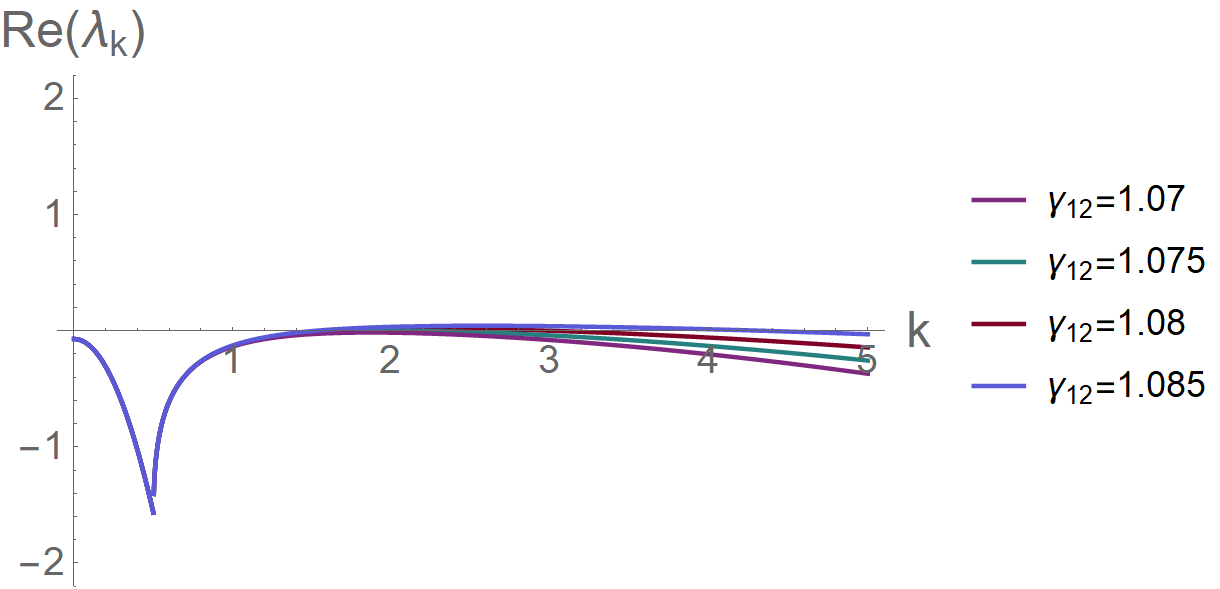}
\includegraphics[width=0.48\textwidth]{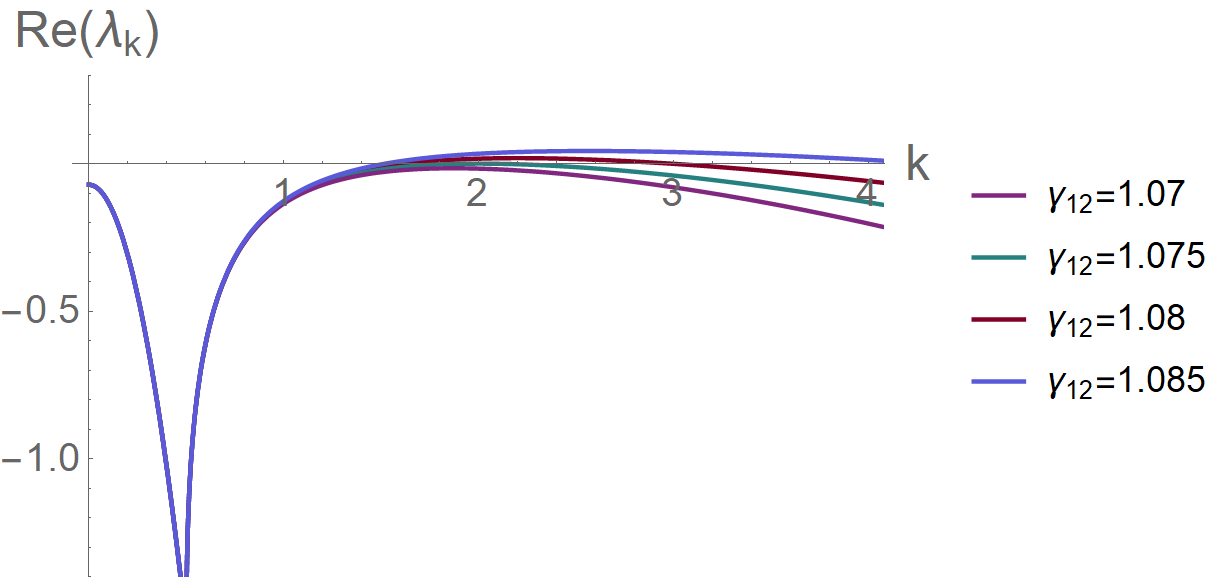}
\caption{\label{fig:aut}
Left panel: plots of the real part of the eigenvalue $\lambda$, solution of (\ref{eqcarc}), as a function of the
wavenumber $k$ for different values of the bifurcation parameter $\gamma_{12}$; in the right panel, a detail of the
same plots. Here again  $\gamma_{11} =3,\,\gamma_{21} =11,\,\gamma_{22} =4, $ and other parameter values as in Table \ref{tab:par}}
\end{figure}

In order to evaluate the effect of cross diffusion, we have assigned a constant value to many of model parameters (as reported in Table 1), we fixed values to $\gamma_{11},\gamma_{21}, \gamma_{22}$ and assumed $\gamma_{12}$ as a bifurcation parameter. 
According to \eqref{gammacr}, it is possible to determine $\gamma_{12}^{cr}$ as the minimum value for Turing instability to occur. Fig. \ref{fig:ic} represents the plots of $h'(k^2)$ as defined in $(\ref{inv})_2$ for different values of the bifurcation parameter $\gamma_{12}$. In this specific example, we have assumed $\gamma_{11}=3,  \gamma_{21}=11  , \gamma_{22}=4$ so that it is $\gamma_{12}^{cr}\approx 1.0748.$ In addition from $(\ref{60})_4$ it is possible to estimate an upper value $\gamma_{12}^{up}\approx 1.09\bar{09}$ above which the condition $(\ref{60})_4$ is no
longer satisfied and  from $(\ref{60})_5$ a lower value $\gamma_{12}^{low}\approx 1.0748$ 
above which $(\ref{60})_5$  is satisfied. 
In the right panel of the same figure, a zoom of the
same plots is shown. As can be seen, for $\gamma_{12} < 1.0748$  the curve does not intersect the horizontal axis, so that there are not unstable modes. As $\gamma_{12}$  increases, the range of unstable modes increases as
well. Similarly, as the bifurcation parameter increases, the real part of the corresponding eigenvalue
becomes positive (see Fig. \ref{fig:aut}).

\begin{figure}
\includegraphics[width=0.48\textwidth]{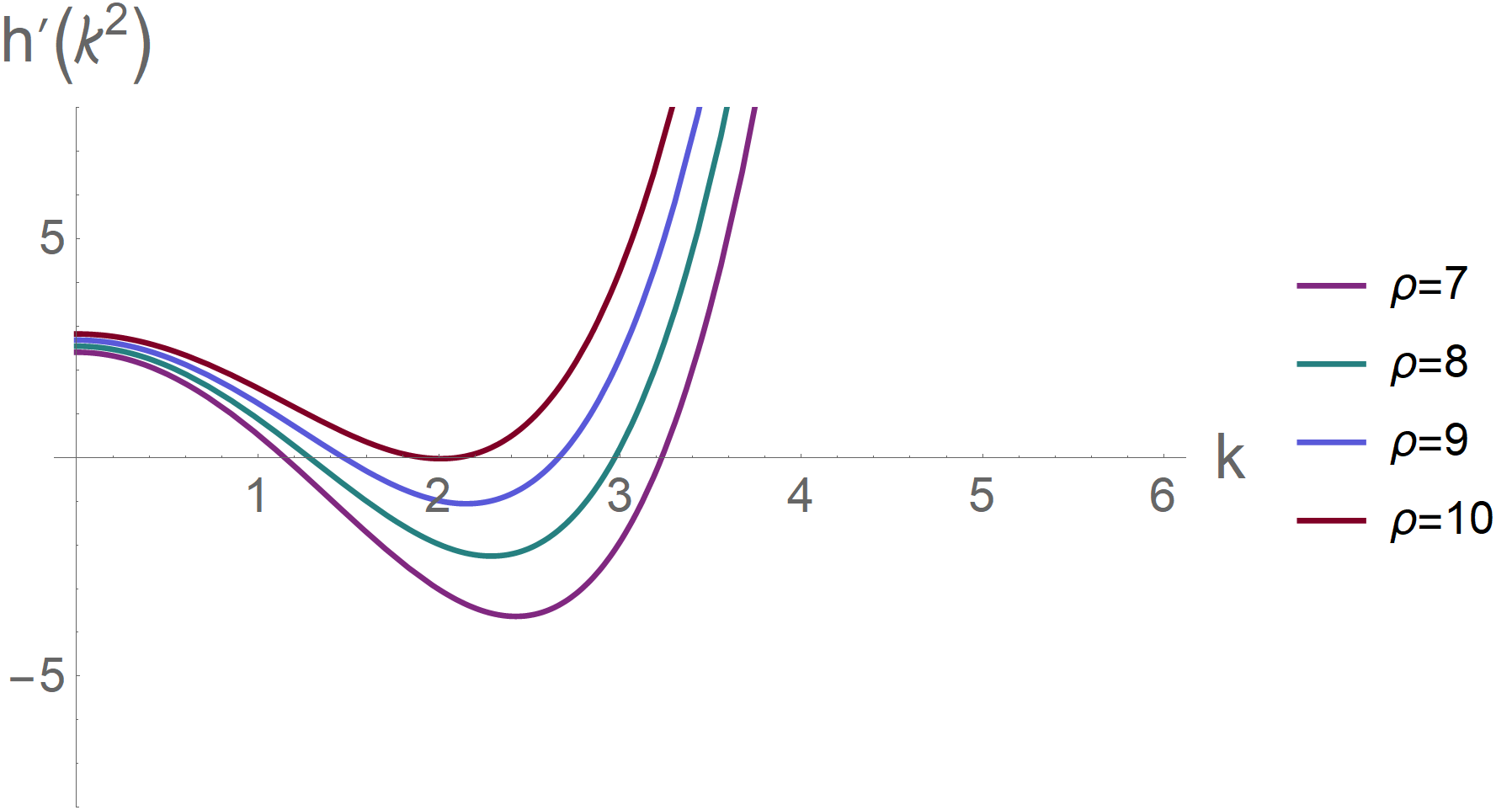}
\includegraphics[width=0.48\textwidth]{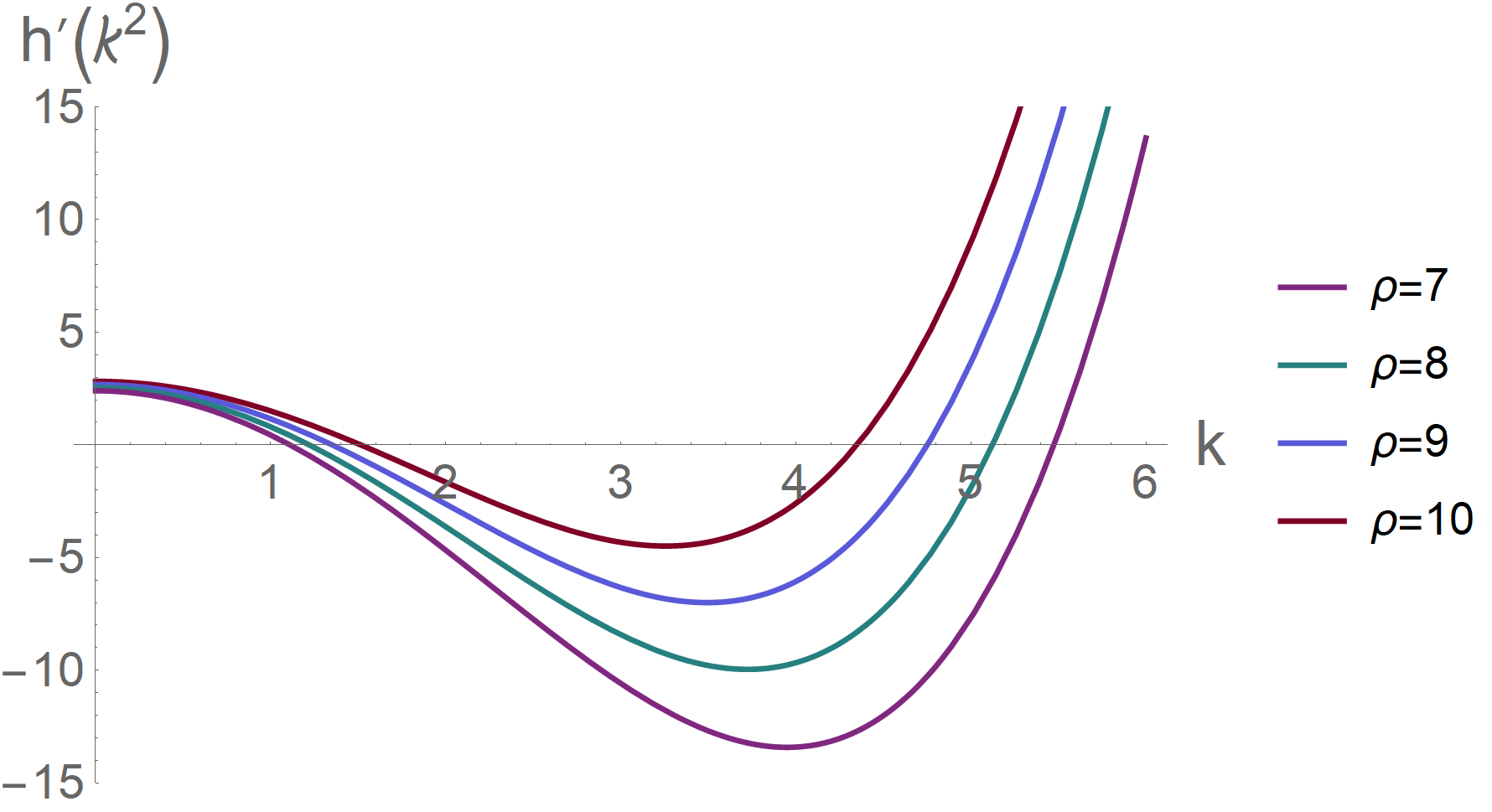}
\caption{\label{fig:fear}
Plots of $h'(k^2)$ as a function of the wavenumber $k$ for different values of the fear level $\rho$ and $\gamma_{11}=3;$ $\gamma_{22}=4$, $\gamma_{21}=11$.  In the left panel, $\gamma_{12} =1.075;$ in the right panel $\gamma_{12}=1.085$.  The other parameter values as in Table \ref{tab:par} }
\end{figure}

We also investigate the effect of the fear level and prey refuge on the unstable modes. As shown in Fig. \ref{fig:fear}, once
the parameter $\gamma_{12}$ is fixed we can notice that higher values of the fear level $\rho$ lead to smaller regions of
unstable modes.  For this reason, as it will be shown in the following experiments, the main effects of a lower fear level are to accelerate the insurgence of patterns and to increase the instability of the system, when the chosen  $\gamma_{12}$ is quite far from its value $\gamma_{12}^{low}.$

\begin{figure}
\includegraphics[width=0.48\textwidth]{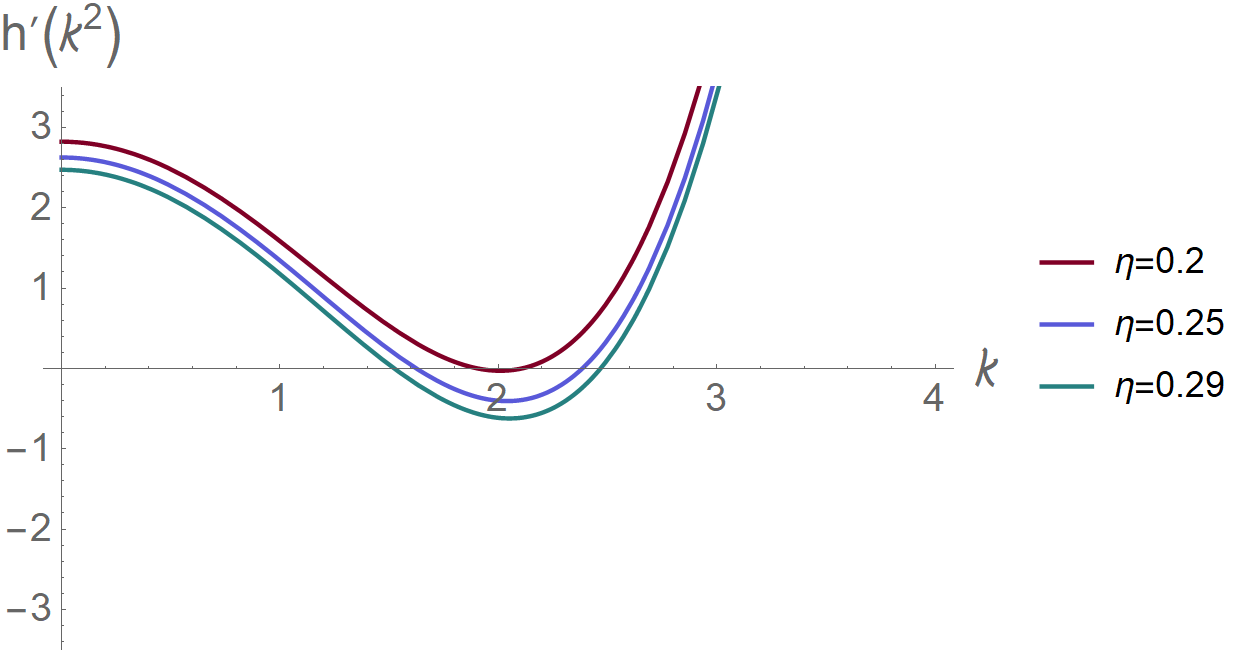}
\includegraphics[width=0.48\textwidth]{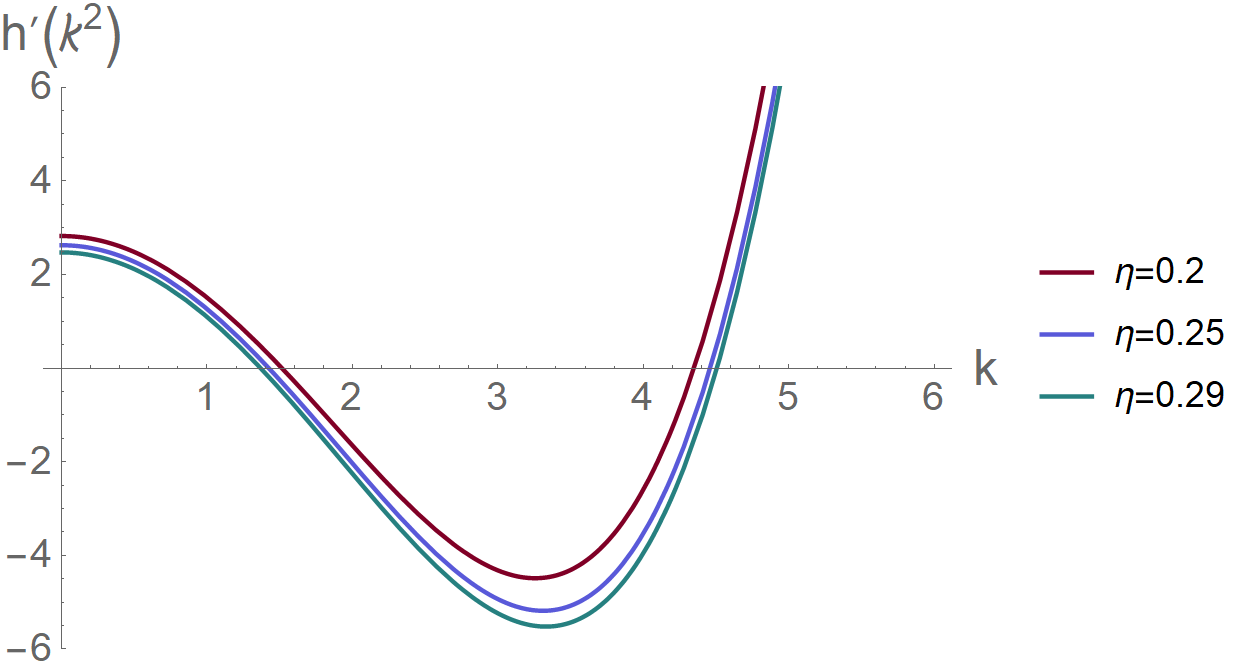}
\caption{\label{fig:refuge}
Plots of $h'(k^2)$ as a function of the wavenumber $k$ for different values of the prey refuge level $\eta$ and  $\gamma_{11}=3;$ $\gamma_{22}=4$, $\gamma_{21}=11$.  In the left panel, $\gamma_{12} =1.075;$ in the right panel $\gamma_{12}=1.085$. The other parameter values as in Table \ref{tab:par}}
\end{figure}

In Fig. \ref{fig:refuge} it has been shown that for fixed values of $\gamma_{12}$ it can be noticed that higher values of the prey refuge  $\eta$ lead to larger regions of unstable modes. In addition, higher values of the prey refuge level imply the increase of the instability and the acceleration of the insurgence of patterns when the chosen  $\gamma_{12}$ is quite far from its value $\gamma_{12}^{low}$.

\begin{figure}
\begin{center}
\includegraphics[width=0.35\textwidth]{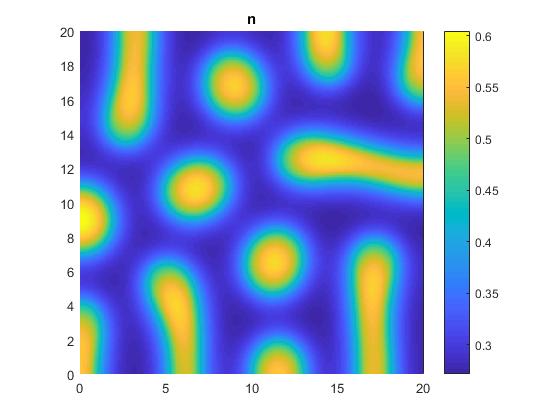}\!\!\!\!\!\!\!\!\!\!
\includegraphics[width=0.35\textwidth]{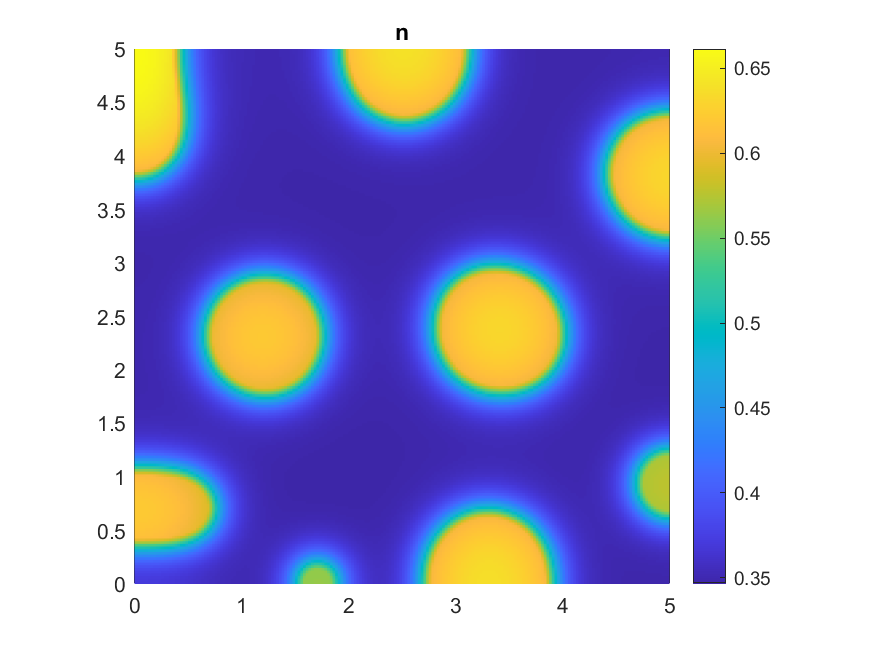}\!\!\!\!\!\!\!\!\!\!
\includegraphics[width=0.35\textwidth]{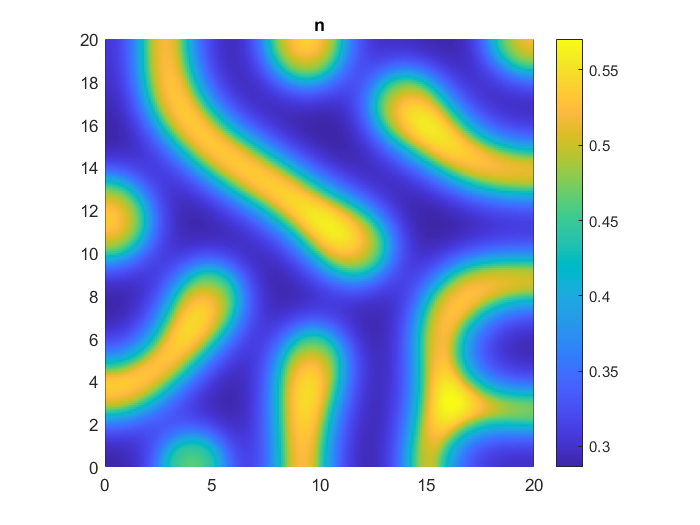}\\
\includegraphics[width=0.35\textwidth]{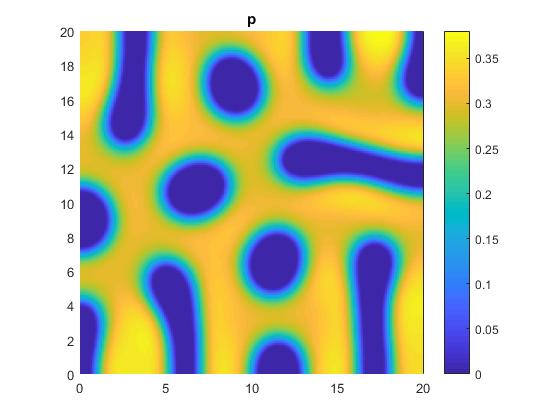}\!\!\!\!\!\!\!\!\!\!
\includegraphics[width=0.35\textwidth]{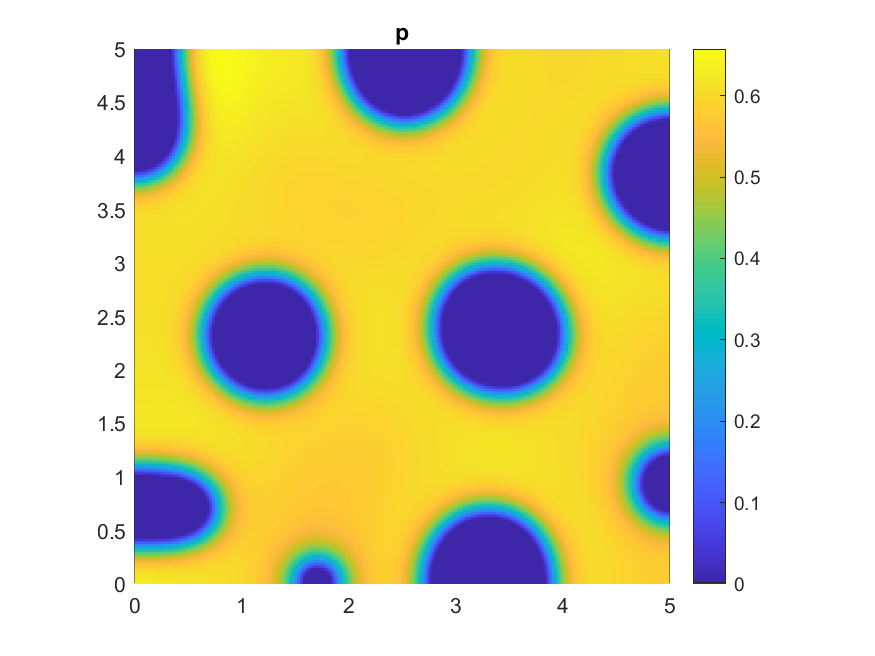}\!\!\!\!\!\!\!\!\!\!
\includegraphics[width=0.35\textwidth]{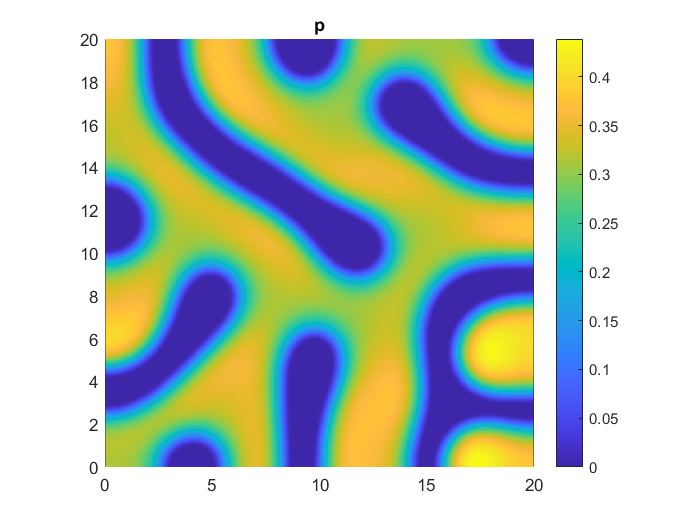}\\
\caption{\label{fig:cross}
Snapshot of pattern formation for different values of diffusion coefficients and $\rho=10$. First column: prey and predator population distribution for $\mu=2, 
\gamma_{11}=3, 
\gamma_{22}=4,  \gamma_{21}=13,  \gamma_{12}=0.6915.$ Second column: prey and predator population distribution for $\mu=3, 
\gamma_{11}=3, 
\gamma_{22}=4,  \gamma_{21}=11,  \gamma_{12}=1.08.$  Third column: prey and predator population distribution for $\mu=2, 
\gamma_{11}=4, 
\gamma_{22}=4,  \gamma_{21}=15.5,  \gamma_{12}=0.7735.$ 
All the other parameter values as in Table 1.
}
\end{center}
\end{figure}

\begin{figure}
\begin{center}
\includegraphics[width=0.46\textwidth]{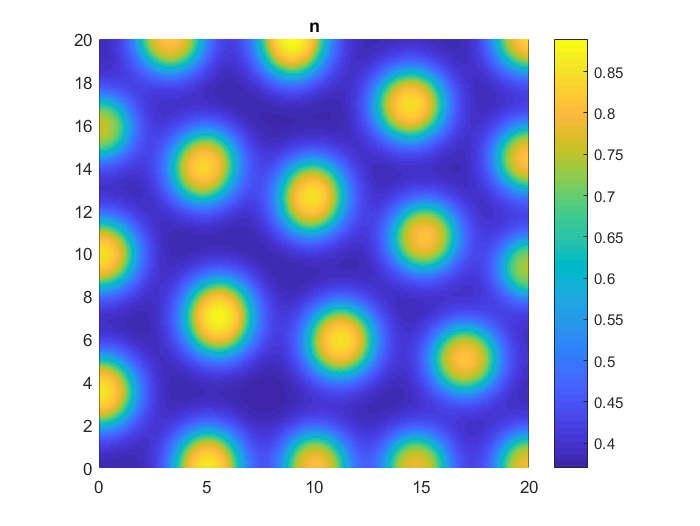}
\includegraphics[width=0.46\textwidth]{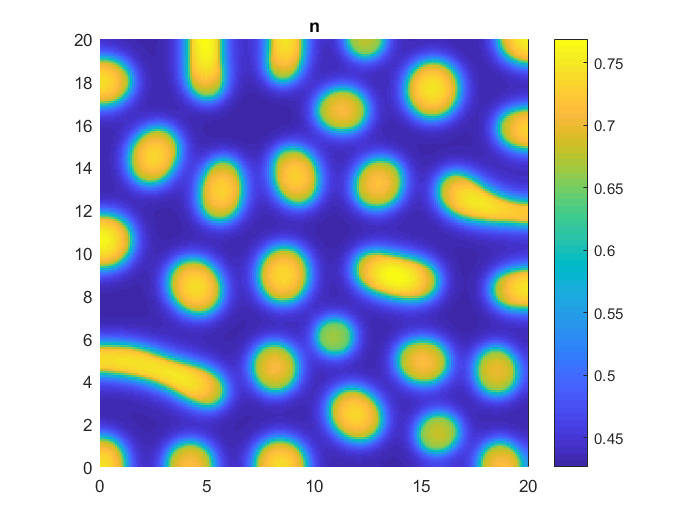}
\includegraphics[width=0.46\textwidth]{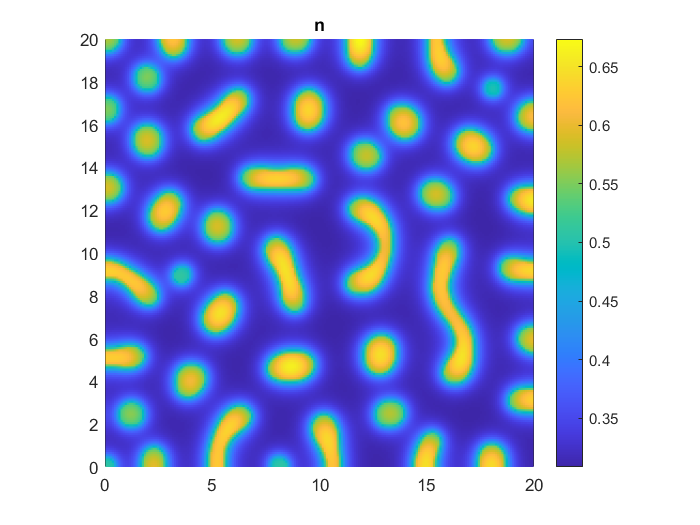}
\includegraphics[width=0.46\textwidth]{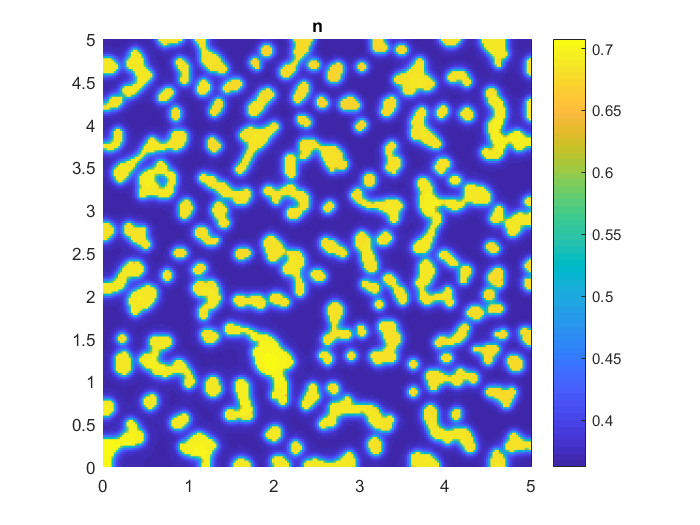}
\caption{\label{fig:fear_varie} Snapshot of pattern formation for prey for different values of fear and diffusion coefficients. First row, from left to right: ($\rho=5, \eta=0.1, \gamma_{11}=4, \gamma_{12}=0.77, \gamma_{22}=3, \gamma_{21}=15.5 $);  ($\rho=5, \eta=0.1, \gamma_{11}=6, \gamma_{12}=1.11, \gamma_{22}=3, \gamma_{21}=15.5$). Second row, from left to right: ($\rho=15, \eta=0.4, \gamma_{11}=2, \gamma_{12}=0.65, \gamma_{22}=3.5, \gamma_{21}=15.5$);   ($\rho=18, \eta=0.5, \gamma_{11}=2, \gamma_{12}=0.99, \gamma_{22}=5, \gamma_{21}=10$).  $\mu=3$ and other parameter values as in Table 1. }
\end{center}
\end{figure}

In the rest of this section we perform some numerical simulations for system \eqref{5.7***} on a two dimensional spatial domain, 
illustrating the final stable patterns for the prey and predator population for different values of diffusion coefficients.
 The numerical simulations are performed by using,
 for the spatial discretization, the  finite difference method with step $\Delta h$ = 0.025 (for a domain $[0,5]\times[0,5]$) and $\Delta h$ = 0.1 (for a domain $[0,20]\times[0,20]$) while for the time
discretization, the explicit Euler’s method, with time step $\Delta t=10^{-6}.$ 
For different values of diffusion coefficients satisfying the Turing conditions, we have obtained different types of Turing patterns representing the distribution of prey and predator species. In every pattern the blue color corresponds to the low density of species and the yellow color corresponds to the high density of species. By various numerical simulations we have observed that prey are distributed generating predominantly spot patterns.
Biologically, yellow spots on the blue background represent that the prey population disposes in isolated regions with high density, moved by fear to better defend themselves from predation. Precisely, as shown in the columns of Fig. \ref{fig:cross}, representing from up to bottom the prey and predator distribution respectively,  hexagonal structures (corresponding to spots), spots and stripes, for prey, appear.  The numerical simulations are well consistent with the theoretical analysis results of amplitude equations.  
\\
Figure \ref{fig:fear_varie} depicts stable patterns of prey distribution, emerging by choosing the value of $\rho$ and $\eta$ and consequently the value of diffusion coefficients (satisfying \eqref{60}) representing the  following scenario. Precisely, the first 
and second 
images represent the distribution of prey  population, with a low level of both fear and refuge ($\rho=5, \eta=0.1$), which  is therefore more tempted to spread in the domain, in search of food and consequently subject to high predatory pressure  ($\gamma_{11}=4, \gamma_{12}=0.77, \gamma_{21}=15.57, \gamma_{22}=3$ and  $\gamma_{11}=6, \gamma_{12}=1.11, \gamma_{21}=15.57, \gamma_{22}=3$ respectively). The third and forth images represent the prey distribution characterized by a higher level of fear and refuge ($\rho=15,\, \eta=0.4$ and $\rho=18,\, \eta=0.5$ respectively) which is therefore less likely to spread to the environment ($\gamma_{11}=2,  \gamma_{12}=0.65, \gamma_{22}=3.5, \gamma_{21}=15.5$  and  $\gamma_{11}=2, \gamma_{12}=0.99, \gamma_{22}=5, \gamma_{21}=10$).
\section{Conclusions}
In this paper, a generalized Leslie-Gower model is introduced to describe the interaction between prey and predator populations. In particular, a random movement of both the species is allowed: at the first, a simple self diffusion is considered for both the species and, after, the more general case in which the diffusion of one species depends on the movement of the other species, is analyzed (cross-diffusion system). A qualitative analysis concerning the boundedness of solutions, existence of absorbing sets in the phase space, the non-existence of non constant steady state, is performed. The linear instability analysis of the coexistence equilibrium (when it exists) is investigated. In particular, conditions guaranteeing self-diffusion and cross-diffusion induced instability, have been determined. Numerical simulations on the obtained results are shown.  
In particular, by varying the values of the model parameters, Turing patterns emerged, representing a spatial redistribution of population in the environment. These results may have wide applications in ecology, biological control for the coexistence of the species in the ecosystem.

\section*{Declarations}

\begin{itemize}
\item {\textbf{Funding}} No funding has been received for this article
\item \textbf{Conflict of interest} We declare we have no competing interest
\item \textbf{Ethics approval} Not applicable
\item \textbf{Consent to participate} Not applicable
\item \textbf{Consent for publication} All authors gave final approval for publication and
agree to be held accountable for the work performed therein
\item {\textbf{Availability of data and materials}}  This article has no additional data
\item \textbf{Code availability} Not applicable 
\item {\textbf{Authors' contributions}} The authors have equally contributed to each part of this paper. They conceived and
thoroughly discussed the main ideas, mathematical models and results of this paper by mutual consent. All
of them carried out in detail the proofs and calculations
\end{itemize}

\section*{Acknowledgments} This paper has been performed under the auspices of
the GNFM of INdAM.

\bibliographystyle{unsrt}
\bibliography{Turing_patterns_Leslie-Gower_predator_prey}

\end{document}